\documentstyle[12pt]{article}

\begin{document}

\baselineskip=22pt plus 1pt minus 1pt

\begin{center}{\large \bf
Deformed harmonic oscillators for metal clusters: Analytic properties 
and supershells}
\bigskip\bigskip

{Dennis Bonatsos$^{\#}$\footnote{e-mail: bonat@mail.demokritos.gr},
D. Lenis$^{\#}$, 
P. P. Raychev$^\dagger$\footnote{e-mail: raychev@inrne.bas.bg},
and P. A. Terziev$^\dagger$\footnote{e-mail: terziev@inrne.bas.bg}
\bigskip

{$^{\#}$ Institute of Nuclear Physics, N.C.S.R.
``Demokritos''}

{GR-15310 Aghia Paraskevi, Attiki, Greece}

{$^\dagger$ Institute for Nuclear Research and Nuclear Energy, Bulgarian
Academy of Sciences }

{72 Tzarigrad Road, BG-1784 Sofia, Bulgaria}}

\end{center}

\bigskip\bigskip
\centerline{\bf Abstract} \medskip

The analytic properties of Nilsson's Modified Oscillator (MO),
which was first introduced in nuclear structure, and of the recently 
introduced, based on quantum algebraic techniques, 
3-dimensional $q$-deformed harmonic oscillator (3-dim $q$-HO)
with u$_q$(3) $\supset$ so$_q$(3) symmetry, which 
is known to reproduce correctly in terms of only one parameter 
the magic numbers of alkali clusters up to 1500 
(the expected limit of validity for theories based 
on the filling of electronic shells), are considered. Exact expressions 
for the total energy of closed shells are determined 
and compared among them. Furthermore, the systematics of the
appearance of supershells 
in the spectra of the two oscillators is considered, showing that the 
3-dim $q$-HO correctly predicts the first supershell closure 
in alkali clusters without use of any extra parameter. 


\newpage

{\bf 1. Introduction} 

Supershells, which are seen as beats of the deviation of the total energy 
of many particle systems from the function describing its average 
behavior vs. the number of particles $N$, 
are known to be a general property of the spectrum of potentials
having sharp edges \cite{Brack}. Supershells in metal clusters have first 
been studied by Nishioka, Hansen and Mottelson \cite{Nishioka} in terms 
of phenomenological mean field potentials. 

In the other hand, using recently developed quantum algebraic techniques
\cite{PPNP},
it has been shown \cite{PRA62}
that the magic numbers appearing in alkali clusters 
can be successfully reproduced up to 1500 (which is the expected limit 
of validity of theories based on the filling of electronic shells
\cite{Martin})
by the 3-dimensional $q$-deformed harmonic oscillator (3-dim $q$-HO), 
which possesses the u$_q$(3) $\supset$ so$_q$(3) symmetry \cite{Terziev}. 
Furthermore, 
the magic numbers appearing in several divalent (Zn, Cd) and trivalent
(Al, In) metal clusters have been satisfactorily reproduced 
\cite{PRA62} by the same 
model, in terms of only one free parameter, the deformation parameter 
$\tau$ (with $q=e^\tau$, where $\tau$ is a real number). It is therefore 
of interest to examine if the 3-dim $q$-HO can predict supershells 
and which these predictions are. It should be noticed that the
calculation of supershells in the framework of the 3-dim $q$-HO will be 
parameter free, since the single parameter of the model has been fixed 
in reproducing the magic numbers for each kind of clusters \cite{PRA62}.  

In addition to the 3-dim $q$-HO, Nilsson's Modified Oscillator (MO)
\cite{Nilsson1,Nilsson2}, 
which has first been used in describing the structure of atomic nuclei,
has also been early employed in describing atomic clusters \cite{Clem}
(after dropping the spin--orbit interaction, which plays an essential role
in nuclear structure but is absent in the case of atomic clusters). 
It is therefore of interest to study the possible appearance of supershells 
in the framework of this model as well.    

For the determination of supershells the method of Ref. \cite{Nishioka} 
can be employed. Before doing so, one has however to examine the analytic 
properties of the spectra of the two oscillators, in order to be able 
to apply meaningful truncation schemes. Furthermore, the average behavior 
of the total energy of a system of many particles (an atomic cluster 
in the present case) as a function of the particle number $N$ should 
be determined, since it is needed in the procedure of the study of 
supershells. As a result, these tasks will be carried out for both models,
before any attempt for the determination of supershells is made. 

Supershells have been early predicted by Balian and Bloch \cite{Balian}
in the study of electrons moving in a spherical cavity, which by analogy 
can be used for the valence electrons in a metal cluster. The comparison 
of the stringent predictions of the theory of Balian and Bloch for various 
characteristics of the supershells \cite{Brack} 
to the results of the present models  
turns out to be a fruitful testing procedure. 

In Section 2 the analytic properties of Nilsson's Modified Oscillator 
will be considered, while the corresponding properties of the 3-dim 
$q$-HO will be studied in Section 3. In Section 4 supershells in Nilsson's 
Modified Oscillator will be studied, while supershells in the framework 
of the 3-dim $q$-HO will be considered in Section 5. Finally in Section 6 
a discussion of the present results and plans for further work will be 
given.  

{\bf 2. Nilsson's Modified Oscillator (MO)} 

The potential of the Modified Oscillator (MO) introduced in nuclear physics 
by Nilsson \cite{Nilsson1,Nilsson2} (with the spin-orbit term omitted) is 
\begin{equation}
V= {1 \over 2} \hbar \omega \rho^2 -\hbar \omega \mu' 
({\bf L}^2 - <{\bf L}^2>_n ), \qquad \rho=r \sqrt {M\omega \over \hbar} ,
\end{equation}
where 
\begin{equation}
<{\bf L}^2>_n = {n(n+3)\over 2}.
\end{equation}
The effect of the ${\bf L}^2$ term is to flatten the bottom of the potential.
In addition it causes an overall compression of the spectrum, which is avoided
through the addition of the  $<{\bf L}^2>_n$ term, which will be 
further discussed later. 

The energy eigenvalues of Nilsson's  modified harmonic oscillator
are \cite{Nilsson1,Nilsson2}
\begin{equation}
E_{nl}= \hbar \omega \left( n+{3\over 2}\right) 
 -\hbar \omega \mu' \left( l(l+1)-{1\over 2} n(n+3)\right), 
\end{equation}
where $n$ is the number of vibrational quanta and $l$ is the 
eigenvalue of the angular momentum, obtaining the values 
$l=n$, $n-2$, \dots, 0 or 1 (depending on $n$ being even or odd, 
respectively). 

The number of states having energy $\left(n+{3\over 2}\right) \hbar \omega$
in the case of even $n$ is 
\begin{equation}
n_1 = \sum^n_{l=0, l = {\rm even}} (2l+1) = {(n+1) (n+2)\over 2},
\end{equation}
where only the even values of $l$ are included in the summation
and the sum 
\begin{equation}
\sum_{x=1}^n x ={n(n+1) \over 2}
\end{equation}
has been used. 
The same result is obtained for odd $n$, in which case only the odd
values of $l$ are included in the summation. Taking into account 
the spin of the particles the number of states having energy 
$\left( n+{3\over 2}\right) \hbar \omega$ is
\begin{equation}
n_2 = (n+1) (n+2).
\end{equation}

The sum of the eigenvalues of ${\bf L}^2$ within each shell in 
the case of even $n$ is 
\begin{equation}
L_1 = \sum_{l=0, l={\rm even}} ^n l(l+1) (2l+1) =
{n(n+1)(n+2)(n+3)\over 4},
\end{equation}
where only the even values of $l$ have been included in the summation
and, in addition to Eq. (5), the sums 
\begin{equation}
\sum_{x=1}^n x^2 = {n(n+1)(2n+1)\over 6}
\end{equation}
and 
\begin{equation}
\sum_{x=1}^n x^3 = {n^2 (n+1)^2\over 4}
\end{equation}
have been taken into account. The same result is obtained for $n$ being odd. 
Taking the spin of the particles into account one has 
\begin{equation}
L_2 = 2 L_1 = {n(n+1)(n+2)(n+3)\over 2}.
\end{equation}
Thus the average per particle of the square of the angular momentum
within each shell is 
\begin{equation}
< {\bf L}^2>_n = {L_1\over n_1} = {L_2 \over n_2} = {n(n+3) \over 2},
\end{equation} 
a result which has already been used in Eqs. (1) and (3), in order to keep
the 
``center of mass'' of each shell constant, i.e. to counterbalance the 
overall compression of the spectrum caused by the ${\bf L}^2$ term alone.   

The total number of particles which can be accommodated in the levels 
of the shells up to the $n$-th shell included, is
\begin{equation}
N=\sum_{x=0}^n (x+1) (x+2) = {(n+1)(n+2)(n+3)\over 3},
\end{equation} 
where the spin of the particles has been taken into account and Eqs.
(5), (6), (8) have been used. It should be remembered throughout 
the present work that $N$ stands for the total number of particles, 
while $n$ stands for the number of vibrational quanta. 

The contribution of the first term of Eq. (3) to the total energy 
of the particles up to the $n$-th shell included (and taking the 
spin of the particles into account) is 
\begin{equation}
E(n)= \hbar \omega \sum _{x=0}^n \left(x+{3\over 2}\right) (x+1) (x+2)  
=\hbar \omega {(n+1) (n+2)^2 (n+3)\over 4},
\end{equation}
where Eq. (6) is used for the degeneracy within each shell and 
Eqs. (5), (8), (9) have been used for performing the summations.
For later use we notice that 
omitting the ground state energy contribution in a similar manner  one finds 
\begin{equation}
E'(n)= \hbar \omega \sum_{x=0}^n x (x+1)(x+2) = 
\hbar \omega {n(n+1)(n+2)(n+3)\over 4},
\end{equation}
while a $n^2$ perturbation in the energy would have given an additional 
term 
\begin{equation}
E_2(n)= \hbar \omega \sum_{x=0}^n x^2 (x+1) (x+2) = 
\hbar \omega {n(n+1)(n+2)(n+3) (4n+1)\over 20}, 
\end{equation}
where the sum
\begin{equation}
\sum_{x=1}^n x^4 = {n(n+1)(2n+1)(3n^2+3n-1)\over 30}
\end{equation}
has been used in addition to Eqs. (5), (6), (8), (9).
 
The contribution of the second term of Eq. (3) to the total energy of the 
particles up to the $n$-th shell included is found by using Eq. (10)
\begin{equation}
E_3(n) = - \hbar \omega \mu' \sum_{x=0}^n {x(x+1)(x+2)(x+3)\over 2},
\end{equation}
while the contribution of the third term of Eq. (3) will be 
\begin{equation}
E_4(n)= +{1\over 2}\hbar \omega \mu' \sum_{x=0}^n x(x+3) (x+1)(x+2),
\end{equation}
where use of Eq. (6) has been made.
We remark that $E_3(n)$ and $E_4(n)$ cancel. Thus we conclude that 
in Nilsson's MO only the first term of Eq. (3) contributes 
to the total energy of the particles up to the $n$-th shell included. 

The average energy per particle (up the $n$-th shell included) is then 
found using Eqs. (12) and (13) to be 
\begin{equation}
<E>= {E(n)\over N} = \hbar \omega \left({3\over 4}n + {3\over 2}\right),
\end{equation}
i.e. the average energy per particle is increasing linearly with the 
shell number, which is the number of vibrational quanta,  
as it is expected for a harmonic oscillator, 
since the angular momentum terms make no contribution, as we have 
already seen. The same result is obtained from Eq. (14)
\begin{equation}
<E'> = {E'(n)\over N} = \hbar \omega {3\over 4} n,
\end{equation}
where the $3/2$ term has been omitted already in Eq. (14), 
while a $n^2$ perturbation in energy, as seen from Eqs. (12) and (15),  
would have given an extra term
\begin{equation}
<E_2(n)>= {E_2(N)\over N} = \hbar \omega {3\over 20} n(4n+1),
\end{equation}
which naturally contains a $n^2$ term. 

The lower part of the spectrum of Nilsson's Modified Oscillator, 
calculated from Eq. (3), is shown in Fig. 1(a) as a function of the 
parameter $\mu'$, together with the magic numbers appearing at different 
parameter values. The following observations can be made: 

a) The magic numbers at the left end of the figure are the ones of the 
3-dimensional isotropic harmonic oscillator. 

b) The magic numbers appearing around $\mu'=0.04$ are in agreement with 
the magic numbers appearing in alkali clusters, up to 138 (see Ref. 
\cite{PRA62}
for more details). The agreement is destroyed beyond this point. 

c) Around the parameter value $\mu'=0.02$ the magic numbers up to 138 
are a mixture of magic numbers of the 3-dimensional isotropic harmonic 
oscillator and magic numbers appearing around $\mu=0.04$ (magic numbers 
of alkali clusters). 

d) Eq. (3) can be rewritten as 
\begin{equation}
E_{nl}= {3\over 2} \hbar \omega +\hbar \omega n \left(1 +{3\mu'\over
2}\right) 
+ \hbar \omega n^2 {\mu'\over 2} -\hbar \omega \mu' l(l+1),
\end{equation}
which clearly shows that in the case of $\mu'>0$ the levels within 
each oscillator shell (characterized by a given value of $n$)
are ordered according to the value of $l$, with the levels with 
higher values of $l$ lying lower in energy, because of the last term 
in Eq. (22). This is indeed the case in Fig. 1(a), although the levels 
have not been labelled by the quantum numbers $n$, $l$ because of lack 
of space. 

e) The level with $l=n$ in particular 
lies lowest in energy within each shell and in general its energy is 
decreasing with increasing $\mu'$, since in this case Eq. (3) takes 
the form
\begin{equation}
E_{nn}= {3\over 2} \hbar \omega+ \hbar \omega n 
\left( 1 +{\mu'\over 2}\right) -\hbar \omega n^2 {\mu'\over 2},
\end{equation} 
its derivative with respect to $\mu'$ being 
\begin{equation}
{dE_{nn}\over d \mu'} = \hbar \omega {1\over 2} n(1-n), 
\end{equation}
which is indeed negative for $n>1$. Indeed the levels which lie lowest 
within each oscillator shell in Fig. 1(a) are the levels with $l=n$, 
which also show negative slope with increasing $\mu'$. 

f) The fact that the $n^2$ term in Eq. (23) appears with a negative sign 
(for $\mu'>0$) can cause difficulties if one tries to describe a system 
with a large number of particles in terms of this oscillator. 
The derivative of $E_{nn}$ with respect to $n$
\begin{equation}
{dE_{nn}\over dn} = \hbar \omega \left(1+{\mu'\over 2} - \mu' n \right) 
\end{equation}
remains positive for 
\begin{equation}
n< {1\over \mu'} +{1\over 2}.
\end{equation}
Beyond this value of $n$ the derivative is negative, meaning that 
levels with higher values of $n$ will lie lower in energy, making it
difficult 
to define a cut-off for the number of shells taken into account. 
For example, if $\mu'=0.04$ (a value which has been found \cite{Clem}
relevant to the description of metal clusters), the derivative remains
positive if $n<25.5$. It is then clear that a reasonable truncation 
of the spectrum is possible only if the number of shells to be taken 
into account is less than 26 (taking the $n=0$ shell into account), 
otherwise no truncation is possible. This drawback of Nilsson's MO 
does not have any consequences in the case of nuclear structure, 
where the model has been first introduced, because of the small
number of particles involved there (for which including the shells
up to $n=8$, shown in Fig. 1(a), suffices), 
but it can cause difficulties if one 
tries to employ this model for the determination of supershells in 
metal clusters, as we shall see later in Sec. 4.  

g) As an extention of f), one sees from Eq. (23) that $E_{nn}$ remains 
positive if $n < {2\over \mu'}+1$. Thus, in the case of $\mu'=0.04$ 
one should have $n<51$, otherwise energies lower than the 
ground state energy will occur. 

{\bf 3. The 3-dimensional $q$-deformed harmonic oscillator (3-dim q-HO)} 

The space of the 3-dimensional $q$-deformed harmonic oscillator consists of 
the completely symmetric irreducible representations of the quantum algebra
u$_q$(3). In this space a deformed angular momentum algebra, so$_q$(3), 
can be defined \cite{Terziev}. 
The Hamiltonian of the 3-dimensional $q$-deformed 
harmonic oscillator is defined so that it satisfies the following 
requirements:

a) It is an so$_q$(3) scalar, i.e. the energy is simultaneously measurable
with the $q$-deformed  angular momentum related to the algebra so$_q$(3) 
and its $z$-projection.   

b) It conserves the number of bosons, in terms of which the quantum 
algebras u$_q$(3) and so$_q$(3) are realized. 

c) In the limit $q\to 1$ it is in agreement with the Hamiltonian of the usual 
3-dimensional harmonic oscillator. 
 
It has been proved \cite{Terziev} that the Hamiltonian of the 3-dimensional 
$q$-deformed harmonic oscillator satisfying the above requirements 
takes the form
\begin{equation}
H_q = \hbar \omega_0 \left\{ [{\cal N}]_q q^{{\cal N}+1} - 
{q(q-q^{-1})\over [2]_q } C_q^{(2)}
\right\},
\end{equation} 
where ${\cal N}$ is the number operator and $C_q^{(2)}$ is the second order 
Casimir operator of the algebra so$_q$(3), while 
\begin{equation}
[x]_q= {q^x-q^{-x} \over q-q^{-1}}
\end{equation} 
is the definition of $q$-numbers and $q$-operators. 

The energy eigenvalues of the 3-dimensional $q$-deformed harmonic oscillator 
are then \cite{Terziev}
\begin{equation}
E_q(n,l)= \hbar \omega_0 \left\{ [n]_q q^{n+1} - {q(q-q^{-1}) \over [2]_q}
[l]_q [l+1]_q \right\}, 
\end{equation}
where $n$ is the number of vibrational quanta and $l$ is the eigenvalue of
the 
angular momentum, obtaining the values
$l=n, n-2, \ldots, 0$ or 1.  

In the limit of $q\to 1$ one obtains ${\rm lim}_{q\to 1} E_q(n,l)=
\hbar \omega_0 n$, which coincides with the classical result. 

For small values of the deformation parameter $\tau$ (where $q=e^{\tau}$)
one can expand Eq. (29) in powers of $\tau$  obtaining \cite{Terziev}
$$
E_q(n,l)= \hbar \omega_0 n -\hbar \omega_0 \tau \left(l(l+1)-n(n+1)\right)
$$
\begin{equation}
-\hbar \omega_0 \tau^2 \left( l(l+1)-{1\over 3} n(n+1)(2n+1) \right)
+ {\cal O} (\tau^3).
\end{equation}

The number of states characterized by a given value of $n$, i.e. the 
number of states in the $n$-th shell, is still given by Eq. (4) if the spin 
is not taken into account, and by Eq. (6) with spin taken into account. 
 
The total number of particles which can be accommodated in the levels 
of all shells up to the $n$-th shell included is still given by 
Eq. (12), with spin taken into account.  

The analogue of the sum of the eigenvalues of ${\bf L}^2$ within each shell 
in the case 
of even $l$ is, in analogy with Eq. (7), given by 
$$
L_{1,q}(n)=\sum_{l=0, l= {\rm even}}^n [l]_q [l+1]_q (2l+1) 
= {1\over (q-q^{-1})^2 (q^2 -q^{-2})^2} 
$$
$$
( (2n+1) (q^{2n+5}+q^{-2n-5})
-(2n+5) (q^{2n+1}+q^{-2n-1}) +5 (q+q^{-1})  -(q^5+q^{-5})
$$
\begin{equation}
-{n(n+3)\over 2} (q^5+q^{-5}+q^3+q^{-3}-2q-2q^{-1})), 
\end{equation} 
where $[l]_q [l+1]_q$ are the eigenvalues of the second order Casimir
operator 
of so$_q$(3), and use of $(2l+1)$ has been made for the degeneracy within a 
shell without taking spin into account. In performing the relevant summations
one needs, in addition to Eq. (5), the sums
\begin{equation}
\sum_{x=0}^n e^{\tau x} = {e^{\tau (n+1)}-1 \over e^\tau -1} ,
\end{equation}
\begin{equation} 
\sum_{x=0}^n x e^{\tau x} = {1\over (e^\tau-1)^2} \left(
n e^{\tau (n+2)} -(n+1) e^{\tau(n+1)} +e^\tau\right),
\end{equation}
of which the first is a simple geometric series, while the second 
can be derived from the first through differentiation with respect to 
the parameter $\tau$. One can easily see that for odd $l$ a result identical 
to the one given in Eq. (31) is obtained. 

Using the definition of the $q$-numbers given in Eq. (28) 
the above result can be rewritten in the form 
$$
L_{1,q}(n)= {q^2\over (q^2-1)^2} 
$$
\begin{equation}
\left( (2n+1) \left[ n+{3\over 2}\right]_{q^2}
-4 \left[ {n+1\over 2}\right]_{q^2} \left[ {n\over 2}\right]_{q^2}
-\left({n\over 2}+1\right) (n+1) \left( \left[ {3\over 2}\right]_{q^2}
+ \left[ {1\over 2}\right]_{q^2} \right) + \left[ {1\over 2} \right]_{q^2}
 \right),
\end{equation}
or equivalently 
\begin{equation}
L_{1,q}(n)= {1\over (q-q^{-1})^2} \left( (2n+1) {[2n+3]_q\over [2]_q} 
-4 {[n]_q\over [2]_q} {[n+1]_q \over [2]_q} -\left({n\over 2}+1\right)
(n+1) [2]_q +{1\over [2]_q} \right),
\end{equation}
where, for example,  use of the identity 
\begin{equation}
[n]_{q^2} = {q^{2n}-q^{-2n} \over q^2-q^{-2} } = {q^{2n}-q^{-2n} \over 
q-q^{-1} }  {q-q^{-1} \over q^2 -q^{-2}} = {[2n]_q \over [2]_q}
\end{equation}
has been repeatedly made. 

Eq. (31) can be rewritten in yet another form by using the definition 
for $Q$-numbers (see, for example, the review article in \cite{PPNP}
for relevant details) 
\begin{equation}
[n]_Q = {Q^n-1\over Q-1},  
\end{equation} 
where 
\begin{equation}
Q=q^2 .
\end{equation}
Using this definition Eq. (31) takes the form
$$
L_{1,q}(n)= {Q^3 \over (Q-1) (Q^2-1)^2} 
$$
$$
\left( [n]_Q Q^{1/2} \left( (Q^2-5)+2n(Q^2-1)\right) 
+ [-n]_Q Q^{-1/2} \left( (Q^{-2}-5)+ 2n (Q^{-2}-1) \right) \right)
$$ 
\begin{equation}
-{ Q^{1/2}\over 2 (Q-1) (Q^2-1)} \left( n^2 (Q+1)^2 -n (Q-1)^2\right) .
\end{equation}

In the limit $q\to 1$,  keeping terms of order up to $\tau^2$ 
one can see that Eq. (34) is reduced to Eq. (7), i.e. it is in agreement 
with the non-deformed case. For this calculation one finds helpful the 
Taylor expansion of $q$-numbers \cite{PPNP}
\begin{equation}
[n]_q = n\pm {\tau^2\over 6} (n-n^3)+{\tau^4\over 360} (7n-10 n^3+3 n^5)
\pm {\tau^6\over 15120} (31 n -49 n^3 +21 n^5 -3 n^7) +\dots,
\end{equation}
where the upper signs correspond to $q$ being a phase factor ($q=e^{i\tau}$
with $\tau$ being real), while the lower signs correspond to $q$ being 
real ($q=e^\tau$ with $\tau$ being real), as in the present case. 

One can now proceed to the calculation of the total energy of the particles 
up to the $n$-th shell included. Using the identity \cite{PPNP}
\begin{equation}
[n]_q q^{n+1} = Q [n]_Q, 
\end{equation}
where $q$-numbers of Eq. (28) ($Q$-numbers of Eq. (37)~) are used in 
the left (right) hand side 
and $Q=q^2$ (Eq. (38)~), 
one finds that the contribution of the first term of Eq. (29)
to the total energy is
$$
E_{1,q}(n) = \hbar \omega _0 \sum_{x=0}^n [x]_q q^{x+1} (x+1) (x+2) 
=\hbar \omega_0 \sum_{x=0}^n Q [x]_Q (x+1) (x+2)  
$$
$$
=\hbar \omega_0  {Q\over (Q-1)^4} \left( (n+1)(n+2) Q^{n+3}
-2 (n+1) (n+3) Q^{n+2} + (n+2) (n+3) Q^{n+1} -2 \right) 
$$
\begin{equation}
-\hbar \omega_0  {Q\over 3 (Q-1)} (n+1)(n+2)(n+3),
\end{equation}  
where, in addition to Eqs. (5), (6), (8), (32), (33) 
one also needs to use the sum
\begin{equation}
\sum_{x=0}^n x^2 e^{\tau x} = {1\over (e^\tau-1)^3} 
\left( n^2 e^{\tau(n+3)}-(2n^2+2n-1) e^{\tau(n+2)} +(n+1)^2 e^{\tau(n+1)}
-e^{2\tau}-e^\tau \right),
\end{equation} 
which is derived from Eq. (33) by differentiation with respect to the 
parameter $\tau$. Using Eq. (37) one can easily see that Eq. (42) can 
be put in the more symmetric form
$$
E_{1,q}(n)= \hbar \omega_0 {Q\over (Q-1)^3} 
$$
$$
\left( (n+1) (n+2) [n+3]_Q -2 
(n+1) [n+2]_Q (n+3) + [n+1]_Q (n+2) (n+3) \right)
$$
\begin{equation}
 -\hbar \omega_0 {Q\over 3(Q-1)} (n+1)(n+2)(n+3).
\end{equation}
In the limit of $Q\to 1$, keeping terms up to $T^3$ 
(where $Q=e^T =q^2=e^{2\tau}$ and thus $T=2\tau$) one finds that 
Eq. (44) agrees with Eq. (13) of the non-deformed case. In this calculation 
it is helpful to use the Taylor expansion of $Q$-numbers \cite{PPNP}
\begin{equation}
[n]_Q= n + {T\over 2} (n^2-n) +{T^2\over 12} (2n^3 -3 n^2+1) +
{T^3\over 24} (n^4-2n^3+n^2) +\dots
\end{equation}

The contribution of the second term of Eq. (29) to the total energy is found
in a similar manner. One has 
$$
E_{2,q}(n)= -\hbar \omega_0 2 {q (q-q^{-1})\over [2]_q}\sum_{x=0}^n
L_{1,q}(x)
=-\hbar \omega_0  {2Q\over (Q^2-1)^3} 
$$
$$
( Q^4 [n]_Q ( 2(Q^2-1)n +(Q^2-2Q-7)) 
-Q^2 [-n]_Q (2(Q^{-2}-1)N+(Q^{-2}-2Q^{-1}-7)) 
$$
\begin{equation}
-{1\over 6} n^2(n+6)  (Q+1) (Q^2-1)^2 +{1\over 6} n (Q+1)  
(Q^4 + 6 Q^3 +34 Q^2+6 Q+1) ),  
\end{equation}
where Eqs. (5), (8), (31), (32), (33)  have been used and the spin 
of the particles has been taken into account. 

The lower part of the spectrum of the 3-dimensional $q$-deformed 
harmonic oscillator, 
calculated from Eq. (29), is shown in Fig. 1(b) as a function of the 
parameter $\tau$, together with the magic numbers appearing at different 
parameter values, while in Fig. 1(c) the full spectrum up to about 1500
particles is exhibited. 
The following comments and comparisons to Nilsson's MO are now in place: 

a) The magic numbers at the left end of both figures are the ones of the 
3-dimensional isotropic harmonic oscillator, as in Fig. 1(a).  

b) The magic numbers appearing around $\tau=0.04$ are in agreement with 
the magic numbers appearing in alkali clusters, up to 1500 (see Ref. 
\cite{PRA62}
for more details), which is the expected limit of validity for theories 
based on the filling of electronic shells \cite{Martin}, while in the case 
of Nilsson's MO the agreement is limited to the magic numbers up to 138
(Fig. 1(a)~).  

c) Around the parameter value $\tau=0.02$ the magic numbers 
up to 138 in Fig. 1(b)  
are a mixture of magic numbers of the 3-dimensional isotropic harmonic 
oscillator and magic numbers appearing around $\tau=0.04$ (magic numbers 
of alkali clusters). Beyond 138 other magic numbers appear. 

d) Numerical calculations show 
that in the case of $\tau>0$ the levels within 
each oscillator shell (characterized by a given value of $n$)
are ordered according to the value of $l$, with the levels with 
higher values of $l$ lying lower in energy, because of the last term in Eq. 
(29). This is indeed the case in Figs. 1(b) and 1(c), although the levels 
have not been labelled by the quantum numbers $n$, $l$ because of lack 
of space. The ordering of the levels within each shell is the same as the 
one appearing in the case of Nilsson's MO (Fig. 1(a)~). 

e) As in Nilsson's MO, the level with $l=n$  
lies lowest in energy within each shell. However, in the present case 
the energy of this level is not decreasing with increasing $\tau$.   
This is true for all levels of the 3-dim $q$-HO: Their energies
increase with increasing $\tau$ (except in the cases of the levels 
with $(n,l)=(0,0)$ and (1,1), the energies of which remain constant
with increasing $\tau$). 

f) As a consequence of e), no difficulties related to truncation appear 
in the present case. Stopping the level scheme at the $l=n$ level 
of a given shell and taking into account all levels with lower $n$ 
(i.e. all levels of the shells lying below the given one),
one makes sure that all levels up to the given level have been included. 
Therefore in the 3-dim $q$-HO reliable truncations can be made, allowing 
for the description of systems with many particles. This point is 
one of the main advantages of the 3-dim $q$-HO in comparison to Nilsson's
MO.  

A few more comments on the comparison between the 3-dim $q$-HO and 
Nilsson's MO are also in place:  

a) The $<{\bf L}^2>_n$ term in Nilsson's Hamiltonian (Eq. (1)) has been put 
in ``by hand'' in order to guarantee that the ``center of mass''
of each shell will remain constant, so that shells will not be 
compressed because of the presence of the ${\bf L}^2$ term. 
In the case of the 3-dim $q$-HO it is clear that the opposite effect
is present: The shells are expanded, because of the extra terms added 
by the $q$-deformation. One way to see this is by comparing the 
first order corrections appearing in Eq. (30) to the last two terms 
in Eq. (3). In both cases the $l(l+1)$ term causes compression of the shells,
while expansion of the shells is caused by the $n(n+1)$ term in the 
case of the 3-dim $q$-HO and by the $n(n+3)/2$ term in the case of 
Nilsson's MO. 
Since the difference of these terms is
\begin{equation}
n(n+1)-{1\over 2} n(n+3) = {1\over 2} n(n-1),
\end{equation} 
which is positive for $n>1$, 
it is clear that the expansion in the case of the 3-dim $q$-HO will be 
stronger than the expansion in the case of Nilsson's MO, which is 
exactly balanced by the compression caused by the $l(l+1)$ term.
As a result, in the case of the 3-dim $q$-HO net expansion of the 
shells will occur, which will be of first order in the parameter $\tau$. 
Comparison of Figs. 1(a) and 1(b) clearly shows this effect. 

b) In Nilsson's MO the last two terms in the energy expression of Eq. (3)
give no net contribution to the total energy up to the $n$-th shell included,
and therefore to the average energy per particle as well, which turns 
out to be proportional to $n$ (see Eq. (19)~). In the 3-dim $q$-HO 
this dependence of the average energy per particle on $n$ is given 
by the lowest order contribution from Eq. (44), while the next order 
contribution from Eq. (44), as well as the lowest order contribution 
from Eq. (46), give terms with higher powers of $n$. This property will 
have to be taken into account when supershells will be considered.  

c) In both models we have derived exact expressions for the total 
energy up to the $n$-th shell included, i.e. for systems of particles 
filling complete shells. In order to consider systems of particles 
for which the last shell is not full, one has to consider numerical methods,
as the Strutinsky method \cite{Strut1,Strut2}, which are beyond the scope 
of the present study. 

{\bf 4. Supershells in Nilsson's Modified Oscillator}

For studying the existence and properties of supershells in Nilsson's 
MO we are going to use the procedure employed by Nishioka {\it et al.}
\cite{Nishioka}. For a given number of particles $N$ the single particle 
energies $E_j(n,l)$ of the $N$ occupied states are summed up
\begin{equation}
E(N)= \sum_{j=1}^N E_j(n,l).
\end{equation} 
This sum is then divided into two parts: A smooth average part $E_{av}$ 
and a shell part $E_{shell}$, which will exhibit the supeshell structure
\begin{equation}
E(N)= E_{av}(N)+E_{shell}(N). 
\end{equation}
For the average part of the total energy a Liquid Drop Model expansion 
is used \cite{Brack}
\begin{equation}
E_{av}(N) = a_1 N^{1/3} + a_2 N^{2/3} + a_3 N. 
\end{equation} 
This expansion should be adequate in the case of Nilsson's MO, for 
which the average energy per particle increases linearly with $N$ 
(see Eq. (19)~), but it should not suffice in the case of the 3-dim
$q$-HO, for which the average energy per particle contains higher order 
terms, as we have seen in the previous section.  
For the latter case an expansion going up to a $N^2$ term 
\begin{equation} 
E_{av}(N)= a_1 N^{1/3} + a_2 N^{2/3} + a_3 N + a_4 N^{4/3} 
+ a_5 N^{5/3} + a_6 N^2
\end{equation} 
should be more appropriate. 
In order to keep the calculations uniform and thus facilitate the comparisons
between the two oscillators, we opted for using the expansion of Eq. (51)
in all cases, although for Nilsson's MO the first three terms would have 
been adequate. Therefore in the case of Nilsson's MO very small 
values will be expected for $a_6$, the coefficient of the $N^2$ term. 

The results for $E_{shell}$ obtained with Nilsson's MO for 8 different values 
of the parameter $\mu'$ (assuming $\hbar\omega=1$) are shown in Fig. 2, 
while the relevant parameter values and rms deviations are exhibited 
in Table 1. As far as the way of calculation is concerned, the following 
comments apply:

a) In all cases the maximum shell used was $n_{max}=30$, implying that 
the last level included in the calculation was the one with $(n,l)=
(30, 30)$, in order to ensure that the complete spectrum up to this point 
has been taken into account, as discussed in Secs. 2 and 3. 

b) From Eq. (26) one sees that for $\mu'=0.02$ one should keep $n<50.5$, 
while for lower values of $\mu'$ the limiting value of $n$ lies 
even higher. This means that including the shells up 
to $n=30$ is a reasonable truncation.  

c) The procedure of the calculation was as follows: First the summations 
described by Eq. (48), resulting in the total energy $E(N)$ for each particle 
number $N$, have been performed. Subsequently, in order to reduce the size 
of the calculation approximately by a factor of 10, the average $E(N)$
was calculated every 11 points (i.e. for $N=6$, 17, 28, \dots), up to 
the point immediately below the cut-off of $(n,l)=(30,30)$, which is reported
in Table 1 as $N_{max}$. These 
averaged values of $E(N)$ were subsequently fitted by the expansion 
of Eq. (51), resulting in the determination of $E_{av}(N)$ at these points. 
Finally $E_{shell}(N)$ has been obtained at these points as the difference 
$E(N)-E_{av}(N)$ and plotted in Fig. 2. 
 
On the contents of Table 1 the following comments can be made: 

a) The behavior of the parameters as functions of $\mu'$ is rather 
smooth, with the exception of $a_6$, the coefficient of $N^2$, 
which assumes very small values, as expected from the comments
following Eq. (51).

b) The maximum number $N_{max}$ of particles below the cut-off is decreasing 
with increasing $\mu'$, as expected from the fact that the level 
$(n,l)=(30,30)$ is getting lower with increasing $\mu'$ (see Eq. (24)~).  

c) The rms deviations are very small, given the fact that the relevant
average 
energies range up to $10^6$. 

On the contents of Fig. 2 the following comments apply: 

a) The gradual development of supershells with increasing $\mu'$ is clearly 
seen in Figs. 2(a)-2(d). While in Fig. 2(a) ($\mu'=0.0001$)
no supershell structure is seen up to $N=10000$, in Fig. 2(b) ($\mu'=0.001$)
the first supershell is seen to be completed around $N=10000$, 
in Fig. 2(c) ($\mu'=0.002$) two supershells are completed up to 
$N=10000$, and in Fig. 2(d) ($\mu'=0.003$)  most of the third supershell
is also completed by $N=10000$. 

b) In Figs. 2(a)-2(g) ($\mu'=0.0001$ - 0.01)
it is clear that $N_{s1}$, the value of $N$ 
at which the first supershell
is completed, is decreasing as a function of $\mu'$. The same is true 
for $N_{s2}$, the value of $N$ at which the second supershell is completed. 
These trends are not followed by Fig. 2(h) ($\mu'=0.02$). 
An explanation of this effect can be gotten from Fig. 1(a),
in which it is clear that for relatively small $\mu'$ ($\mu'<0.01$,
for example) the order of the levels remains the same and only their 
individual energies change, while at $\mu'=0.02$ and beyond the 
order of the levels changes drastically, especially at higher energies. 
This drastic mixing of the levels at $\mu'=0.02$ and beyond can also 
explain the increasing difficulty in determining the closing of 
supershells for large $\mu'$ especially at high energies, 
Fig. 2(h) ($\mu'=0.02$) being the clearest example for this case.  

c) According to the study by Balian and Bloch of electrons moving 
in a spherical cavity \cite{Balian}, which by analogy can be applied to the 
valence electrons in a metal cluster \cite{Brack}, the minima of the 
shell energy, $E_{shell}$, which correspond to shell closures, 
should appear at equidistant positions 
(i.e. they should exhibit a periodicity) as a function of $N^{1/3}$
within each supershell.  
From the data used for plotting Fig. 2 one can see that this condition 
is approximately fulfilled. 

{\bf 5. Supershells in the 3-dimensional $q$-deformed harmonic oscillator} 

The procedure described in the previous section has been used in exactly the 
same way for the determination of supershells in the case of the 3-dim 
$q$-HO. 

The results for $E_{shell}$ obtained with the 3-dim $q$-HO  for 10 different 
values 
of the parameter $\tau$ (assuming $\hbar\omega=1$) are shown in Fig. 3, 
while the relevant parameter values and rms deviations are exhibited 
in Table 2. In order to facilitate comparisons between the two models, 
the first 8 values of $\tau$ used here are the same as the values of 
$\mu'$ used in the previous section. 

As far as the way of calculation is concerned, the maximum shell used 
in the first 8 cases was $n_{max}=30$, as in the previous section, 
in order to facilitate comparisons between the two models. 
Lower $n_{max}$ has been used only in the last two cases, 
as shown in Table 2, in order to keep the rms deviation small. 
 
On the contents of Table 2 the following comments can be made: 

a) The behavior of the parameters as functions of $\tau$ is  
smooth, even in the case of $a_6$, the coefficient of $N^2$,
which in the case of Nilsson's MO was not showing smooth behavior. 
 
b) The maximum number $N_{max}$ of particles below the cut-off is decreasing 
with increasing $\tau$. This can be explained by looking at Fig. 1(c). 
The level used as the cut-off is the level with $(n,l)=(n_{max}, n_{max})$,
which, as explained in Sec. 3,  is the lowest level within the $n_{max}$-th 
shell and, as seen in Fig. 1(c), increases very slowly with increasing 
$\tau$. It is then clear that the higher the value of $\tau$, the more 
will be the levels coming from below and crossing over the cut-off level, 
resulting in the decrease of $N_{max}$ with increasing $\tau$ seen 
in Table 2.    

c) As in the case of Nilsson's MO, the rms deviations are very small, 
given the fact that the relevant average energies range up to $10^6$. 

On the contents of Fig. 3 the following comments apply: 

a) The gradual development of supershells with increasing $\tau$ is clearly 
seen in Figs. 3(a)-3(d) ($\tau=0.0001$ - 0.003), 
which look very similar to Figs. 2(a)-2(d) ($\mu'=0.0001$ - 0.003) 
of the previous section. 

b) In Figs. 2(a)-2(g) ($\tau=0.0001$ - 0.01) 
it is clear that $N_{s1,q}$, the value of $N$ 
at which the first supershell
is completed, is decreasing as a function of $\tau$, almost in the same way 
as $N_{s1}$ is decreasing as a function of $\mu'$ in the case of 
Nilsson's MO, the only difference being that for numerically equal 
values of $\tau$ and $\mu'$ the corresponding value of $N_{s1,q}$ 
is slightly higher that the relevant value of $N_{1s}$, a fact that 
can be explained by the general property of the spectrum of the 3-dim 
$q$-HO to expand more rapidly than the spectrum of Nilsson's MO, 
as seen in Sec. 3.  
The same is true for $N_{s2,q}$, $N_{s3,q}$, $N_{s4,q}$, i.e. the values 
of $N$ at which the second, third, and fourth supershells are completed. 
These trends are not followed by Figs. 3(h) ($\tau=0.02$),
3(i) ($\tau=0.038$), 3(j) ($\tau=0.05$), which correspond to 
higher values of $\tau$, where the mixing of the levels is very strong
in comparison to the picture existing at low values of $\tau$ (i.e.
for $\tau< 0.01$, as seen in Fig. 1(c)~). 

c) Despite the fact that the systematics of supershell closures are modified 
beyond $\tau=0.02$ because of the strong mixing of levels, supershells 
are still seen beyond this point, while this was not possible in the case 
of Nilsson's MO, because of truncation related problems, as we have seen in 
Sec. 3. It should be noticed that this is exactly the region of $\tau$
values which has been found relevant for the description of metal 
clusters \cite{PRA62}. 

d) In the case of alkali clusters, Na in particular, 
the first supershell is known to occur around $N=1000$ 
\cite{Nishioka,Brech,Pedersen,Genzken}, 
which is quite in agreement with what is seen in Fig. 3(i), which corresponds 
to $\tau=0.038$, the parameter value for which the magic numbers of 
alkali clusters are correctly reproduced \cite{PRA62} up to 1500 particles,
which is the expected limit of validity of theories based on the filling
of electronic shells \cite{Martin}. 
It should be noticed that only one free parameter 
exists in the theory, namely $\tau$, which has been fixed in Ref. \cite{PRA62}
in order to reproduce the magic numbers of alkali clusters. Therefore 
no free parameter has been left over in the calculation of supershells.
The fact that the present calculation leads to a reasonable prediction 
of the position of the first supershell closure is a quite stringent 
test of the present theory. 

e) As we have already mentioned, the theory by Balian and Bloch 
\cite{Balian} for electrons moving in a spherical cavity, which by analogy 
can be applied to the valence electrons of metal clusters \cite{Brack}, 
predicts that the minima of the shell energy, $E_{shell}$, 
which correspond to shell closures, should 
appear within each supershell 
at equidistant positions (i.e. they should exhibit a periodicity) 
when plotted versus $N^{1/3}$. 
From the data used for plotting Fig. 3 
one can see that this condition is approximately fulfilled. 

f) A further prediction of the theory by Balian and Bloch \cite{Balian}
is that the plot of the cubic root $N_i^{1/3}$ of the magic numbers $N_i$, 
corresponding to shell closures, versus the index $i$, counting the shells,
should be a straight line \cite{Brack}. 
In the case of alkali clusters, in particular, in which the triangular 
and squared closed orbits are supposed to dominate \cite{Brack,Brech},
the slope of the line should be 0.603~. In order to make a preliminary 
estimate of the degree to which this prediction is fulfilled, one can employ
the data used in plotting Fig. 3(i), which corresponds to $\tau=0.038$, 
i.e. to the parameter value found appropriate \cite{PRA62} for reproducing 
the magic numbers of several alkali clusters. 
 One can then see that this requirement is roughly fulfilled, 
although the slope appears to be gradually decreasing at large $i$, 
an effect which might be due to missing corrections mentioned
in comment h).  

g) In the case of Al clusters, magic numbers have been determined
experimentally in detail up to 1200 electrons \cite{Persson}, while
additional experimental results up to 2700 electrons exist \cite{Lerme}.
The slope of $N_i^{1/3}$ vs. $i$ in this case is considerably lower
(around 0.32) \cite{Brack}, indicating that closed orbits other than the 
triangular and squared ones should be present in the Balian and Bloch 
approach \cite{Lerme}. 
It will be interesting to examine if the 3-dim $q$-HO can provide any
prediction for supershells in Al clusters, after choosing the value 
of the $\tau$ 
parameter in order to reproduce the right slope of $N_i^{1/3}$ vs. $i$. 

h) In order to guarantee
the reliability of such predictions, it is of interest to study in advance
the influence of any modifications imposed by the Strutinsky method
\cite{Strut1,Strut2}, which can be applied in cases in which the last shell
is open, which are beyond the realm of the present analytic study
of Section 3. The influence of the well known quantum mechanical effect that
$\hbar \omega_0$ should be decreasing with increasing number of particles
in the cluster \cite{Nilsson2,Kotsos1,Kotsos2} should also be taken into
account.

{\bf 6. Discussion}

The main results of the present study are in summary the following:

a) The 3-dimensional $q$-deformed harmonic oscillator, 
which is known \cite{PRA62} to describe very well
the magic numbers of alkali clusters up to 1500 particles (the
expected limit of validity for theories based on the filling of electronic
shells), is found in the present study able to produce supershell
structures, succesfully predicting the first supershell in alkali clusters
without involving any free parameter in addition to the deformation
parameter $\tau$, which has been fixed for reproducing the magic numbers.

b) It should be noticed that these successes of the 3-dim $q$-HO are
largely due to terms in the Hamiltonian induced by the symmetry, which
make the spectrum of the 3-dim $q$-HO to expand with increasing
shell number $n$ more rapidly than the corresponding spectrum of Nilsson's
Modified Oscillator, allowing, among other things, for reliable truncations
to be performed.

c) The successful prediction of the magic numbers
can be considered as evidence that the 3-dimensional $q$-deformed
harmonic oscillator owns a symmetry [the u$_q$(3) $\supset$ so$_q$(3)
symmetry, which is a nonlinear deformation of the u(3) symmetry of
the spherical (3-dimensional isotropic) harmonic oscillator]
appropriate for the description of the physical systems under study.
The use of this symmetry for predicting supershells in other kinds
of metal clusters (Al clusters, for example) is an interesting open problem.
The influence of using open shells (taken into account by the Strutinsky
method
\cite{Strut1,Strut2}), as well as of the well known quantum mechanical fact
that $\hbar \omega_0$ should be decreasing with increasing number of particles
in the cluster \cite{Nilsson2,Kotsos1,Kotsos2}, should be taken into account
before any final predictions can be made.

{\bf Acknowledgements}

Support from the Bulgarian Ministry of Science and Education under Contract
Nos. $\Phi$-415 and $\Phi$-547 is gratefully acknowledged.

\newpage


\centerline{\bf Table 1}

\bigskip
{Parameters used for fitting the average part of the total 
energy (see Eq.~(51)) in the case of Nilsson's Modified Oscillator, 
for various values of the model parameter $\mu'$ (see Eq.~(3)),
corresponding to the cases exhibited in Fig. 2. 
The parameters are dimensionless, since we have assumed $\hbar \omega=1$
(see Eq.~(3)) throughout. The highest shell $n_{max}$ and
the number of particles 
$N_{max}$ included in each calculation, as well as the relevant rms deviation 
$\sigma$, are also shown. See Section 4 for further discussion. 
}

\bigskip\bigskip\bigskip

\begin{tabular}{l r r r r r r r r r}
\hline
$\mu'$&$a_1$  & $a_2$& $a_3$& $a_4$& $10^3 a_5$& $10^5 a_6$ 
& $n_{max}$ & $N_{Max}$ & $\sigma$ \\
\hline
0.0001 &13.349&-6.556&-0.218&0.972 & 4.451& 6.853 & 30&10027    &22.06\\
0.001  &-0.025& 0.281&-1.482&1.080 & 0.080&-0.132 & 30&10027    & 7.12\\
0.002  &-0.022& 0.181&-1.439&1.073 & 0.489& 1.053 & 30& 9609    & 4.19\\
0.003  &-1.281& 1.080&-1.651&1.094 &-0.445& 0.282 & 30& 9246    & 3.40\\
0.005  &-1.073& 1.013&-1.659&1.096 &-0.623& 0.043 & 30& 8641    & 2.76\\
0.007  &-0.830& 0.910&-1.654&1.097 &-0.724&-0.626 & 30& 7937    & 4.16\\
0.01   &-5.448& 3.432&-2.155&1.143 &-2.645& 0.383 & 30& 7189    & 4.88\\
0.02   & 5.692&-3.891&-0.376&0.938 & 9.135&-38.541& 30& 4890    & 3.50\\
\hline
\end{tabular}

\newpage


\centerline{\bf Table 2} 
\bigskip

{Parameters used for fitting the average part of the total 
energy (see Eq.~(51)) in the case of the 3-dimensional $q$-deformed 
harmonic oscillator for various values of the model parameter $\tau$ 
(see Eq.~(29), with $q=e^\tau$),
corresponding to the cases exhibited in Fig. 3. 
The parameters are dimensionless, since we have assumed $\hbar \omega_0=1$
(see Eq.~(29)) throughout. The highest shell $n_{max}$ and the 
number of particles 
$N_{max}$ included in each calculation, as well as the relevant rms deviation 
$\sigma$, are also shown. See Section 5 for further discussion. 
}

\bigskip\bigskip\bigskip

\begin{tabular}{l r r r r r r r r r}
\hline
$\tau$&$a_1$  & $a_2$& $a_3$& $a_4$& $10^3 a_5$& $10^5 a_6$ 
& $n_{max}$ & $N_{Max}$ & $\sigma$ \\
\hline
0.0001 &13.585&-6.554&-0.228&0.972 & 4.464 &-6.776 & 30&10027    &22.12\\
0.001  & 0.753&-0.061&-1.424&1.073 & 0.867 &-0.272 & 30&10027    & 7.58\\
0.002  & 1.458&-0.496&-1.318&1.059 & 2.075 &-1.136 & 30& 9686    & 4.68\\
0.003  &-3.358& 2.186&-1.855&1.106 & 0.615 & 2.501 & 30& 9389    & 3.52\\
0.005  & 0.667& 0.166&-1.500&1.074 & 2.791 & 2.422 & 30& 8795    & 3.49\\
0.007  &-2.631& 1.965&-1.867&1.106 & 2.204 & 7.580 & 30& 8421    & 3.88\\
0.01   &-3.657& 2.568&-2.013&1.120 & 2.523 &15.399 & 30& 7893    & 6.75\\
0.02  &-17.409&11.661&-4.267&1.379&-10.052 &88.547 & 30& 7189    & 9.03\\
0.038&-55.417&41.650&-13.067&2.615&-96.648&482.365 & 26& 4648    &17.98\\
0.05&-64.123&55.364&-19.160&3.777&-202.994&999.866 & 22& 3020    &16.63\\
\hline
\end{tabular}

\newpage 

\centerline{\bf Figure captions} 
\bigskip

{\bf Fig. 1} (a) Energy spectrum of Nilsson's Modified Oscillator 
(in units of $\hbar \omega$, see Eq.~(3)) 
as a function of the (dimensionless) model parameter $\mu'$. Magic numbers 
are shown at the main gaps. 
(b) Energy spectrum of the 3-dimensional $q$-deformed harmonic oscillator 
(in units of $\hbar \omega_0$, see Eq.~(29)) 
as a function of the (dimensionless) model parameter $\tau$ (with $q=e^\tau$, 
where $\tau$ is real). 
(c) Same as (b), but extended to higher energy levels. 

\medskip
{\bf Fig. 2}
Shell part ($E_{shell}$) of the total energy (in units of $\hbar \omega$,
see Eq.~(3)) for Nilsson's Modified Oscillator vs. the number of particles
$N$. The values of the (dimensionless) 
parameter $\mu'$ are the same as these listed in Table 1, 
together with the details of the calculation. See Section 4 for further 
discussion. 

\medskip
{\bf Fig. 3} 
Shell part ($E_{shell}$) of the total energy (in units of $\hbar \omega_0$, 
see Eq.~(29)) for the 3-dimensional $q$-deformed harmonic oscillator vs.
the number of particles $N$. The values of the (dimensionless)
parameter $\tau$ are the same
as these listed in Table 2, together with the details of the calculation.
See Section 5 for further discussion.

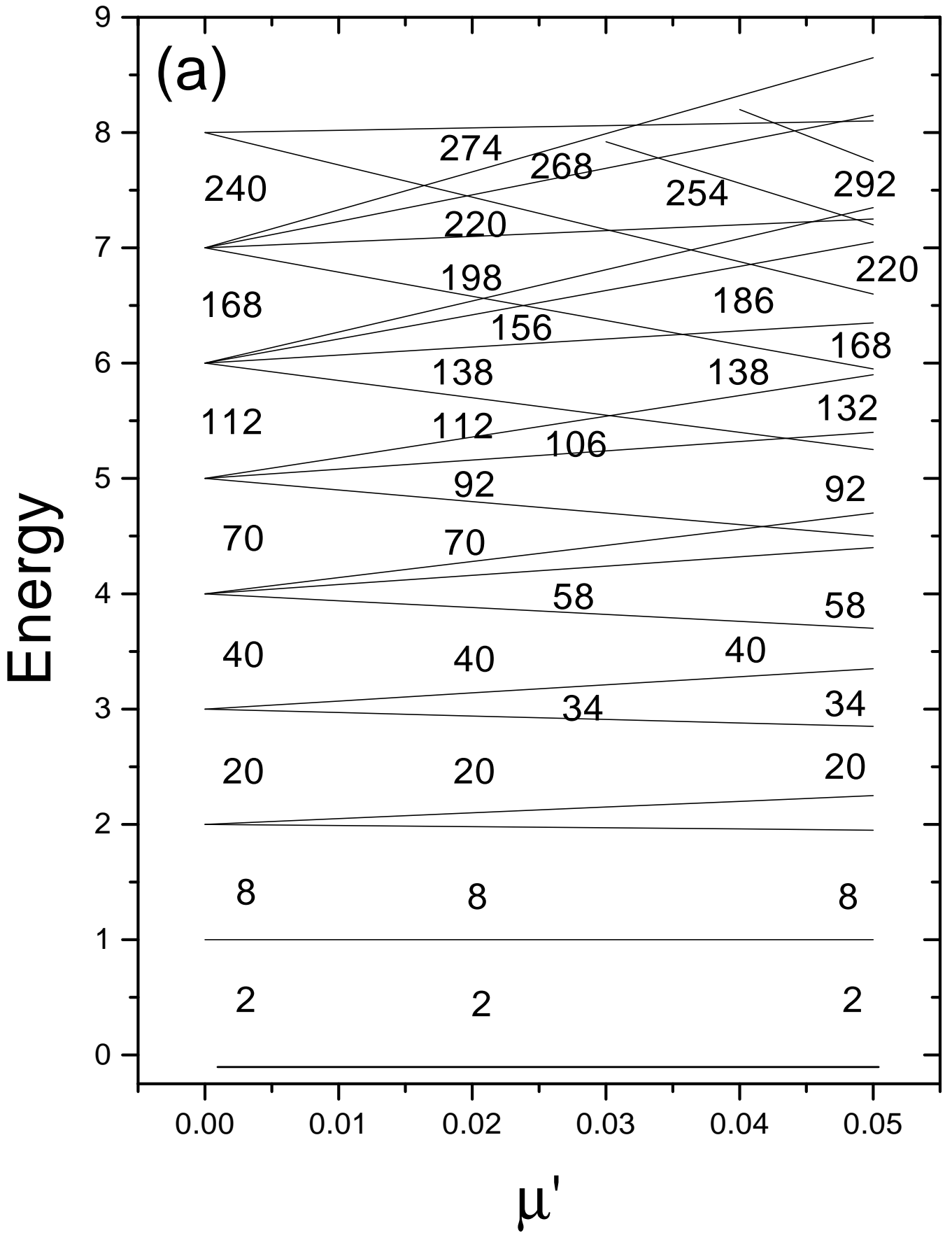
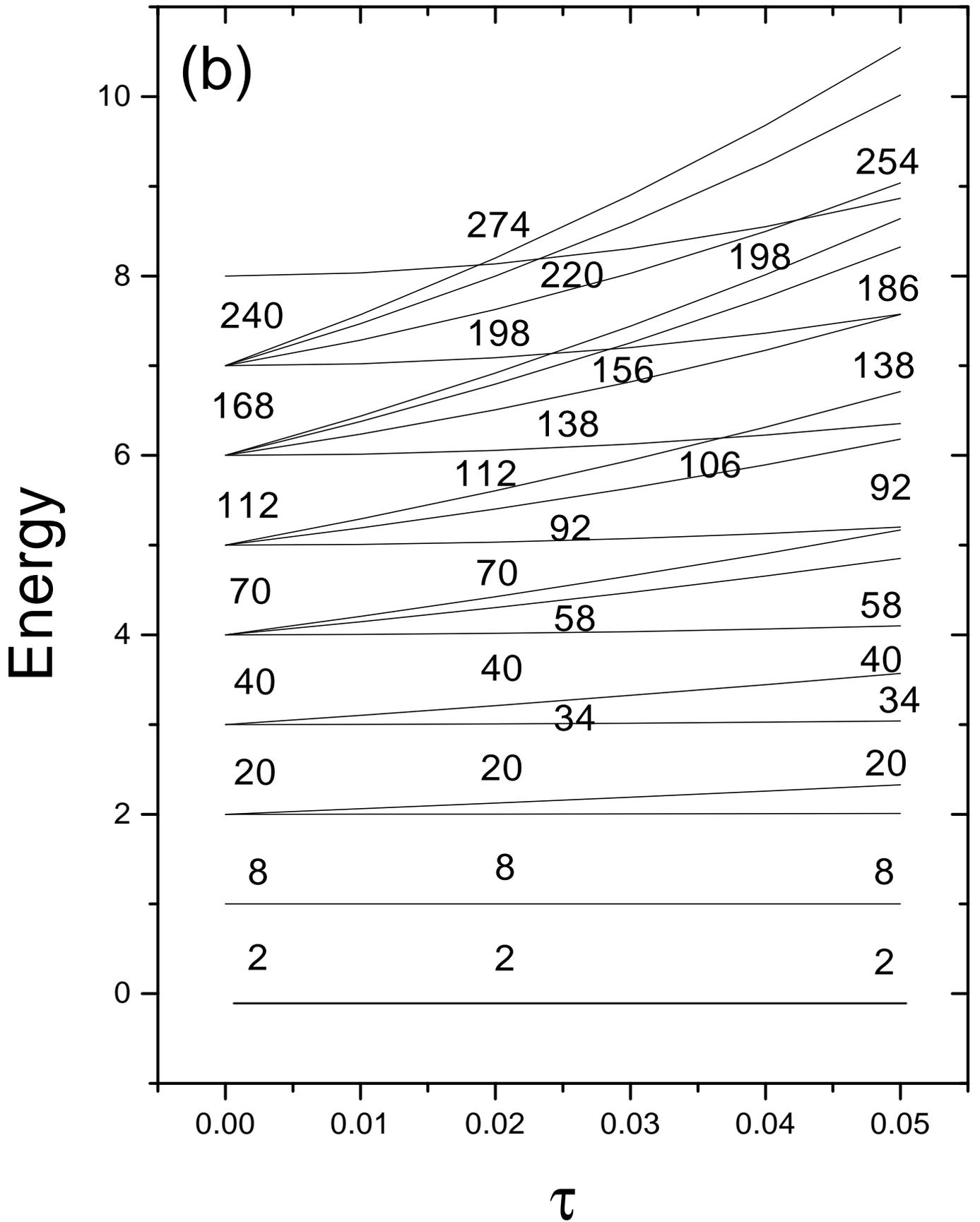
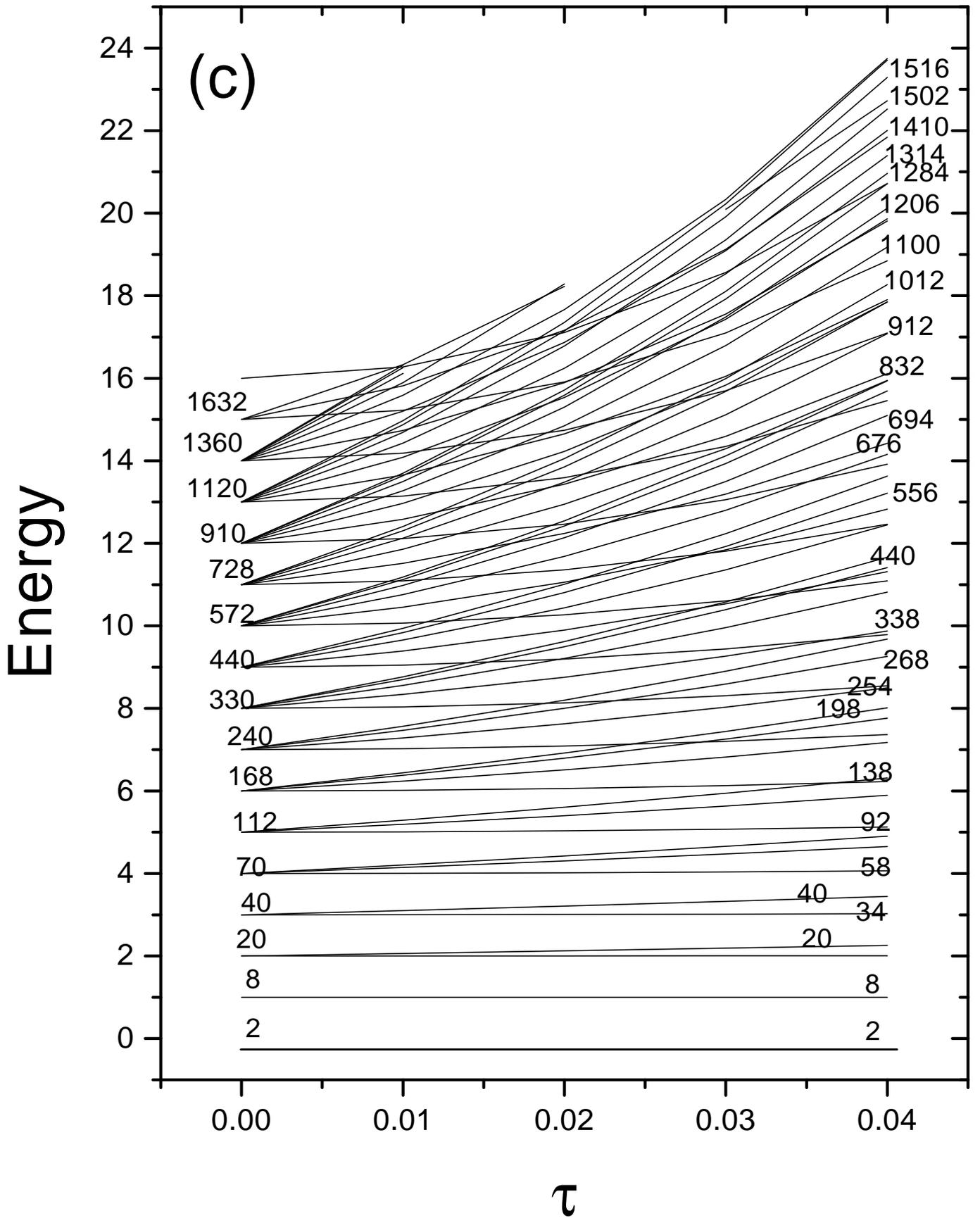
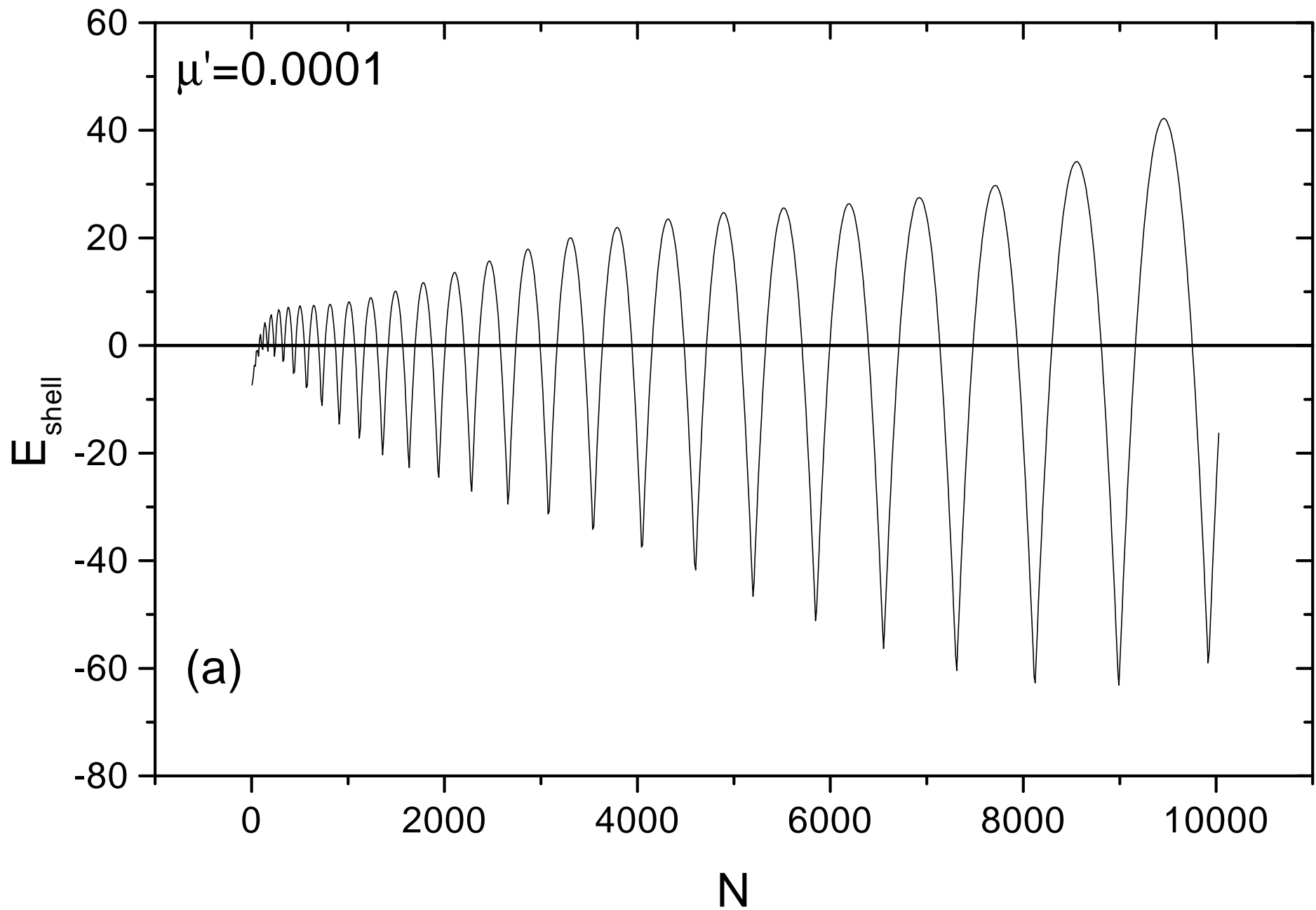
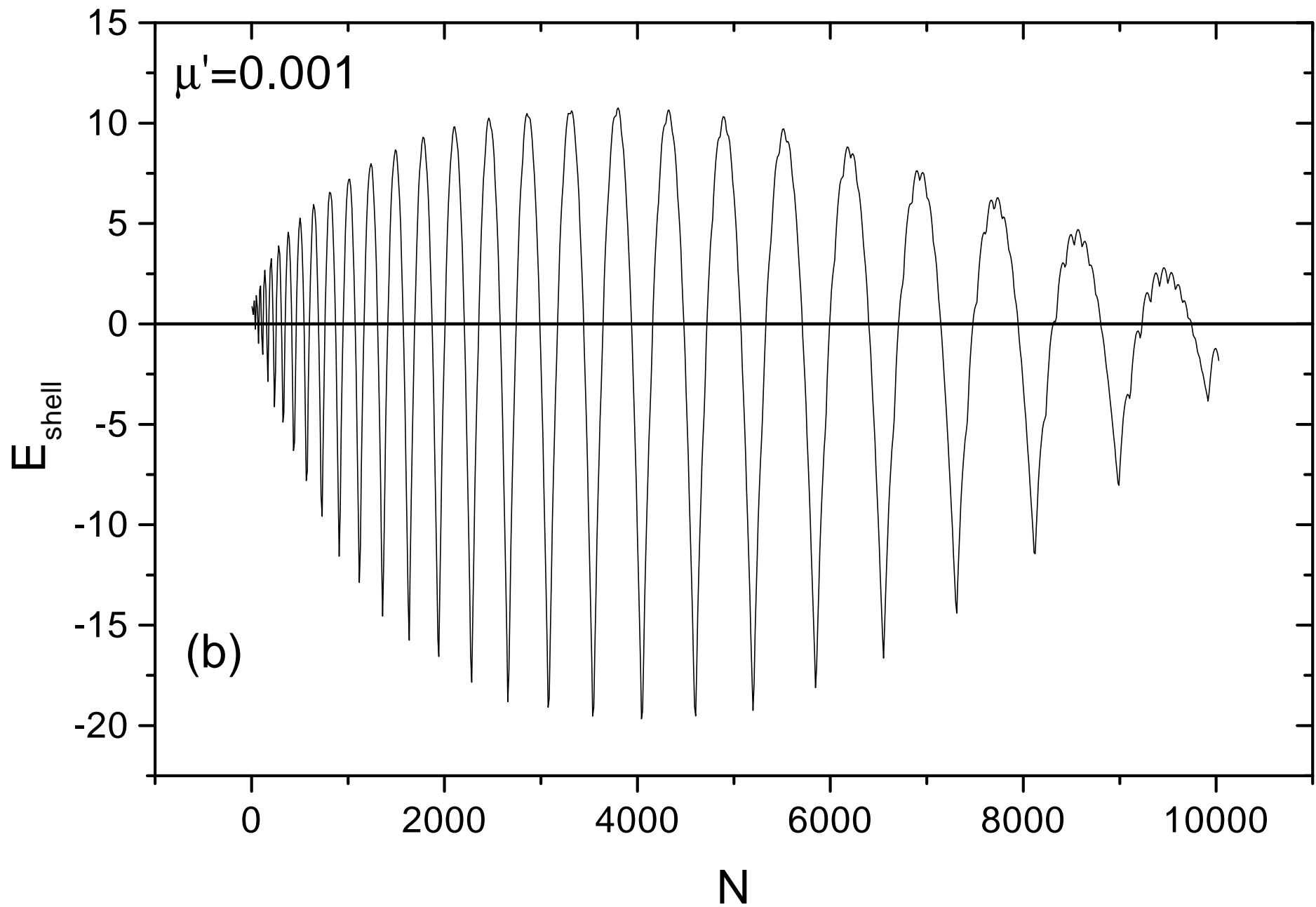
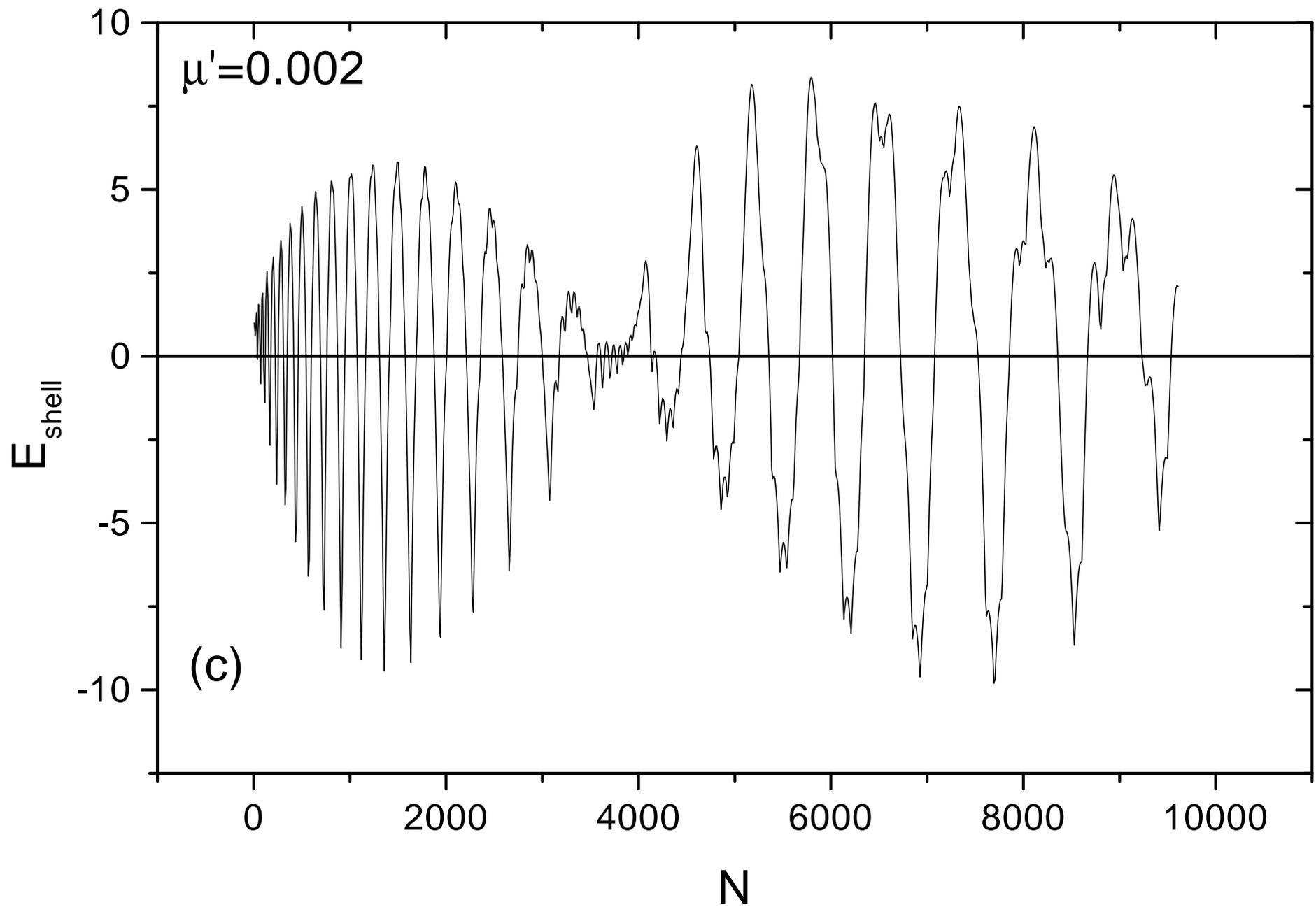
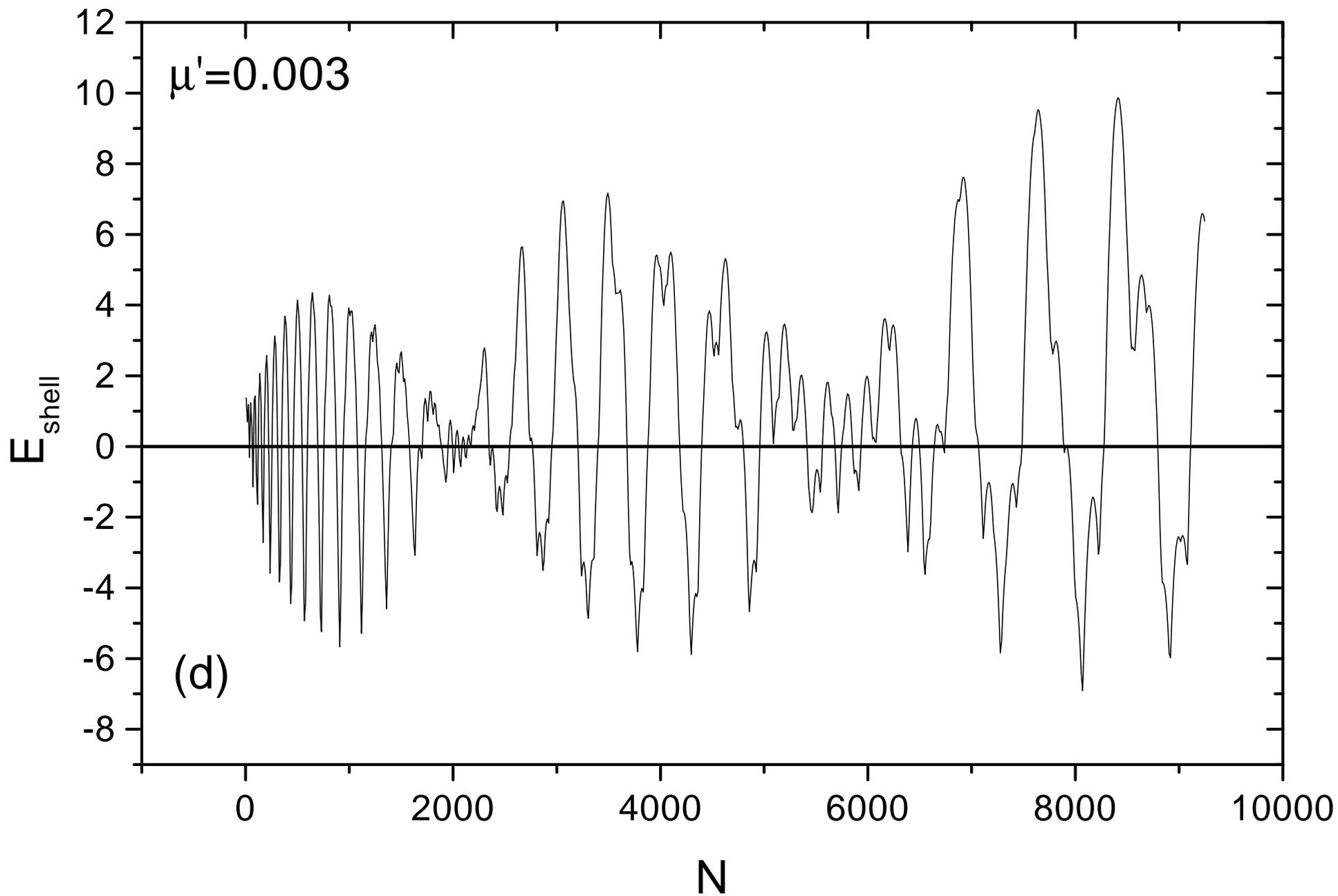
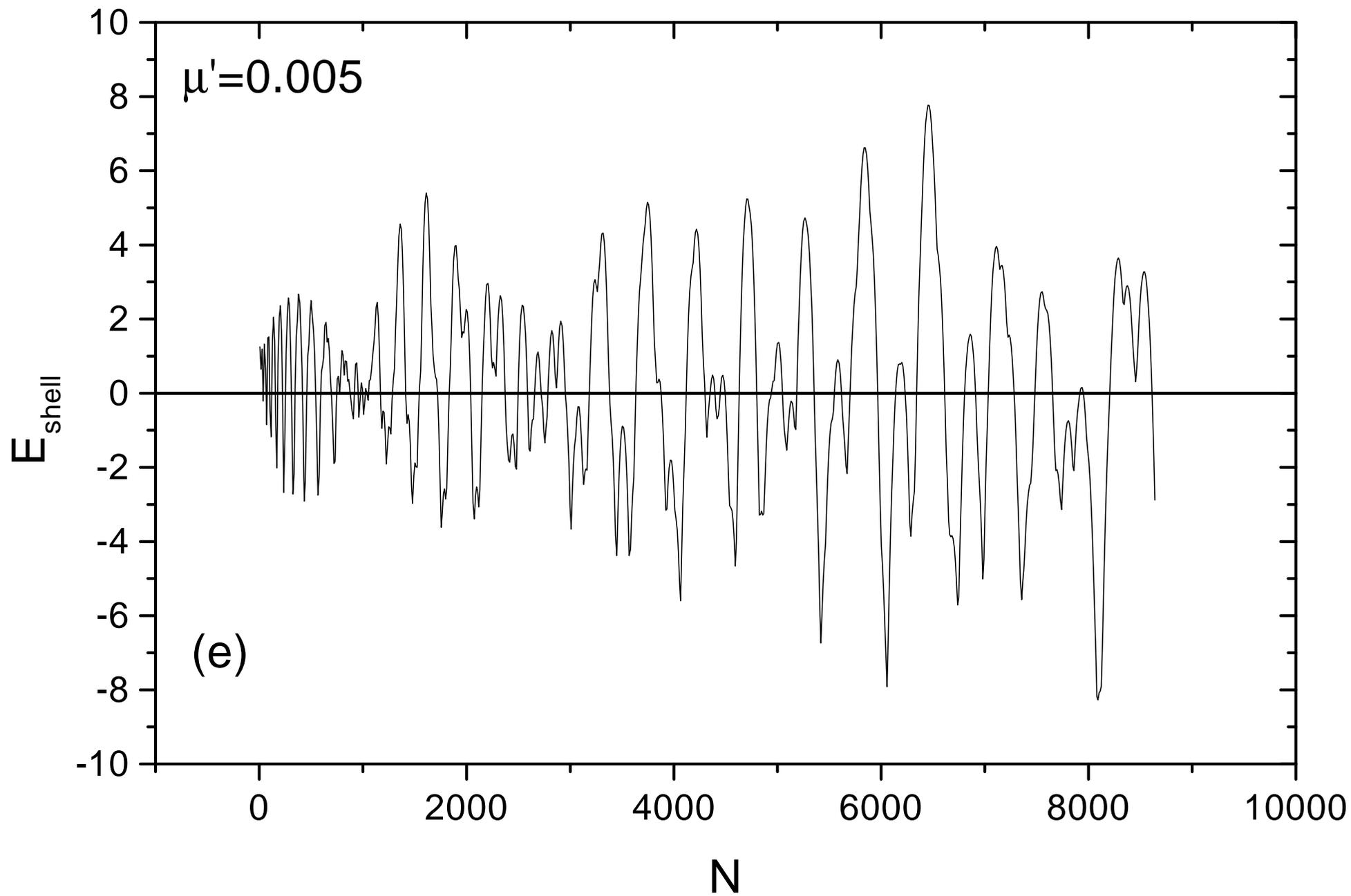
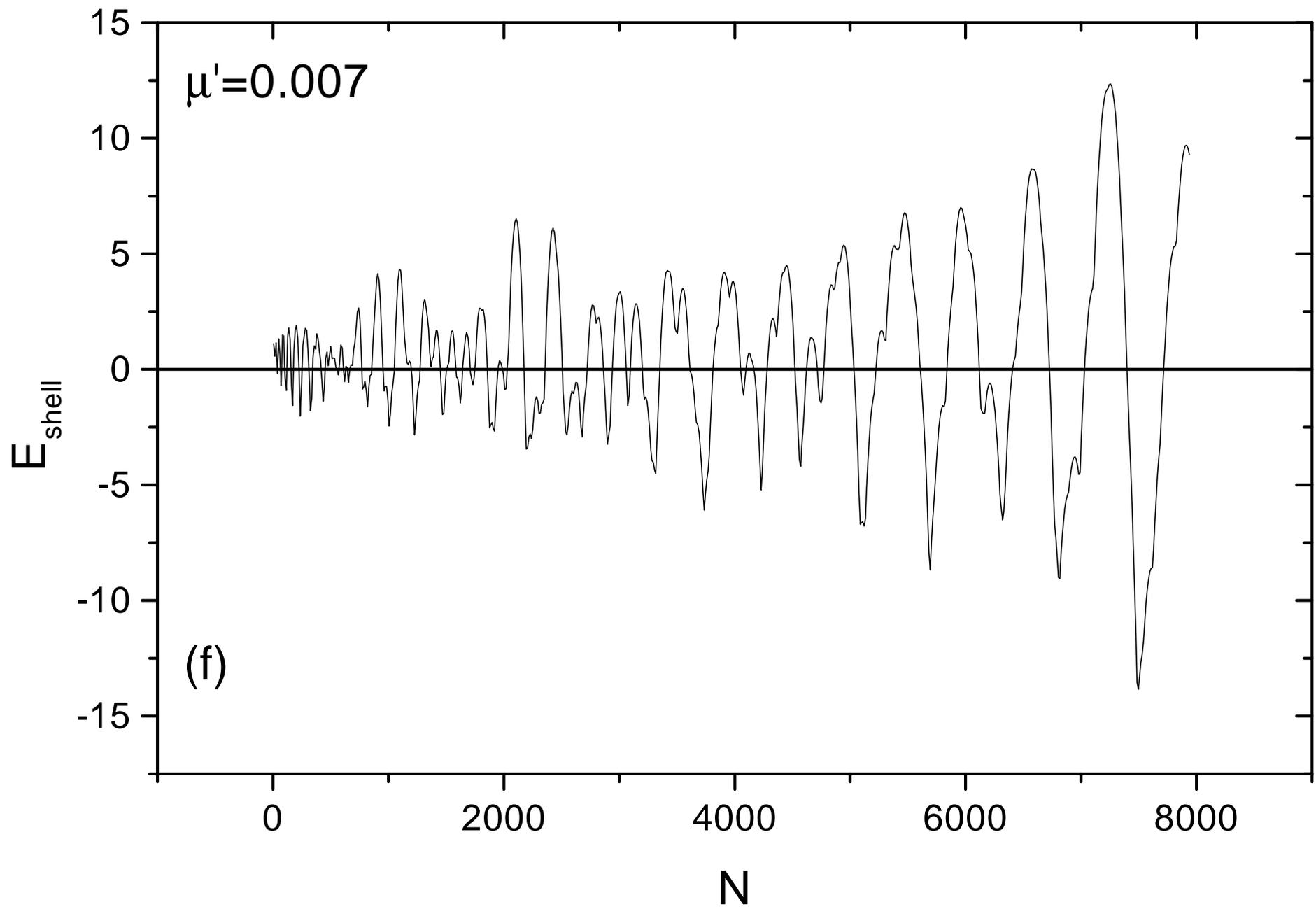
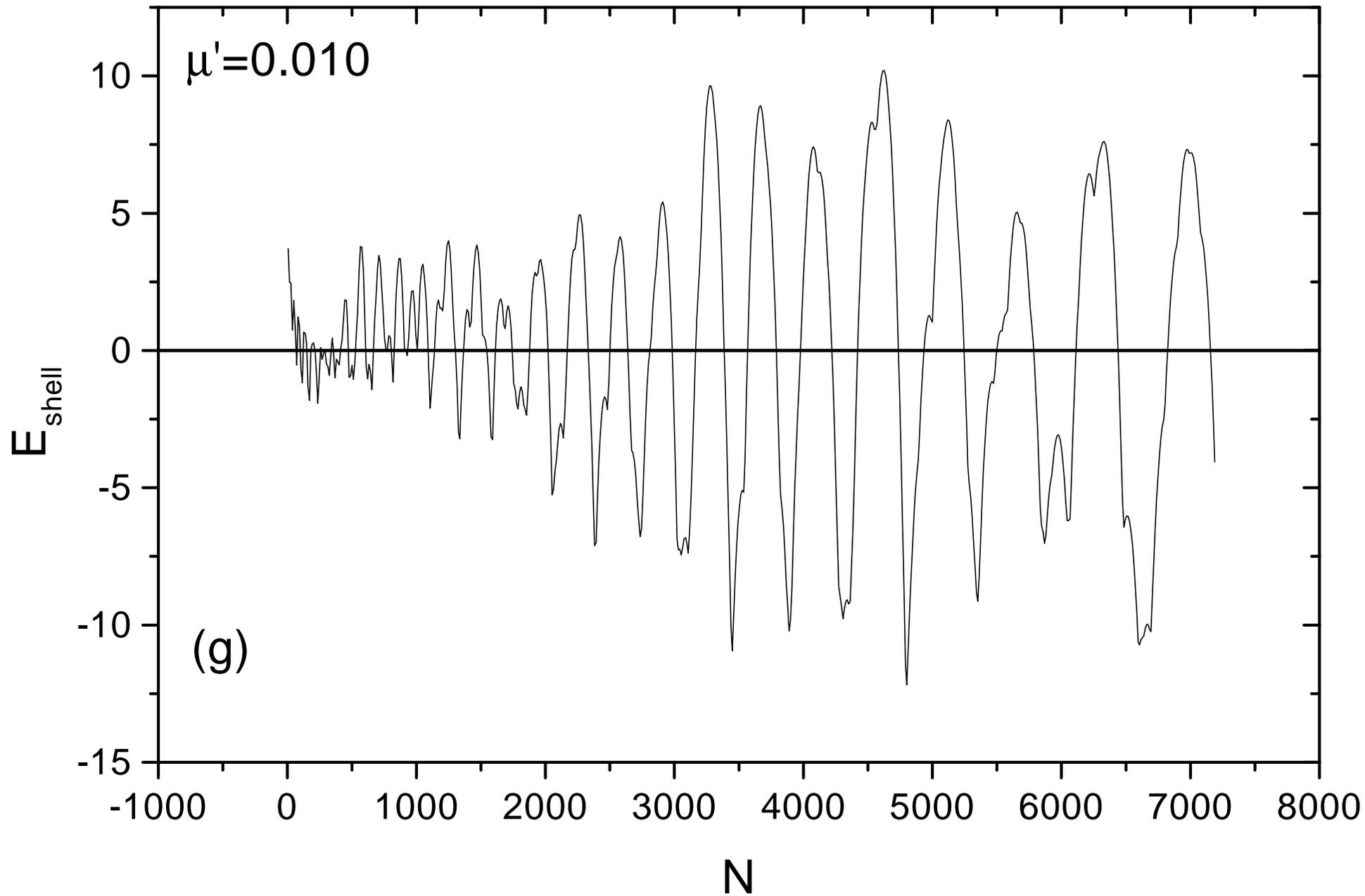
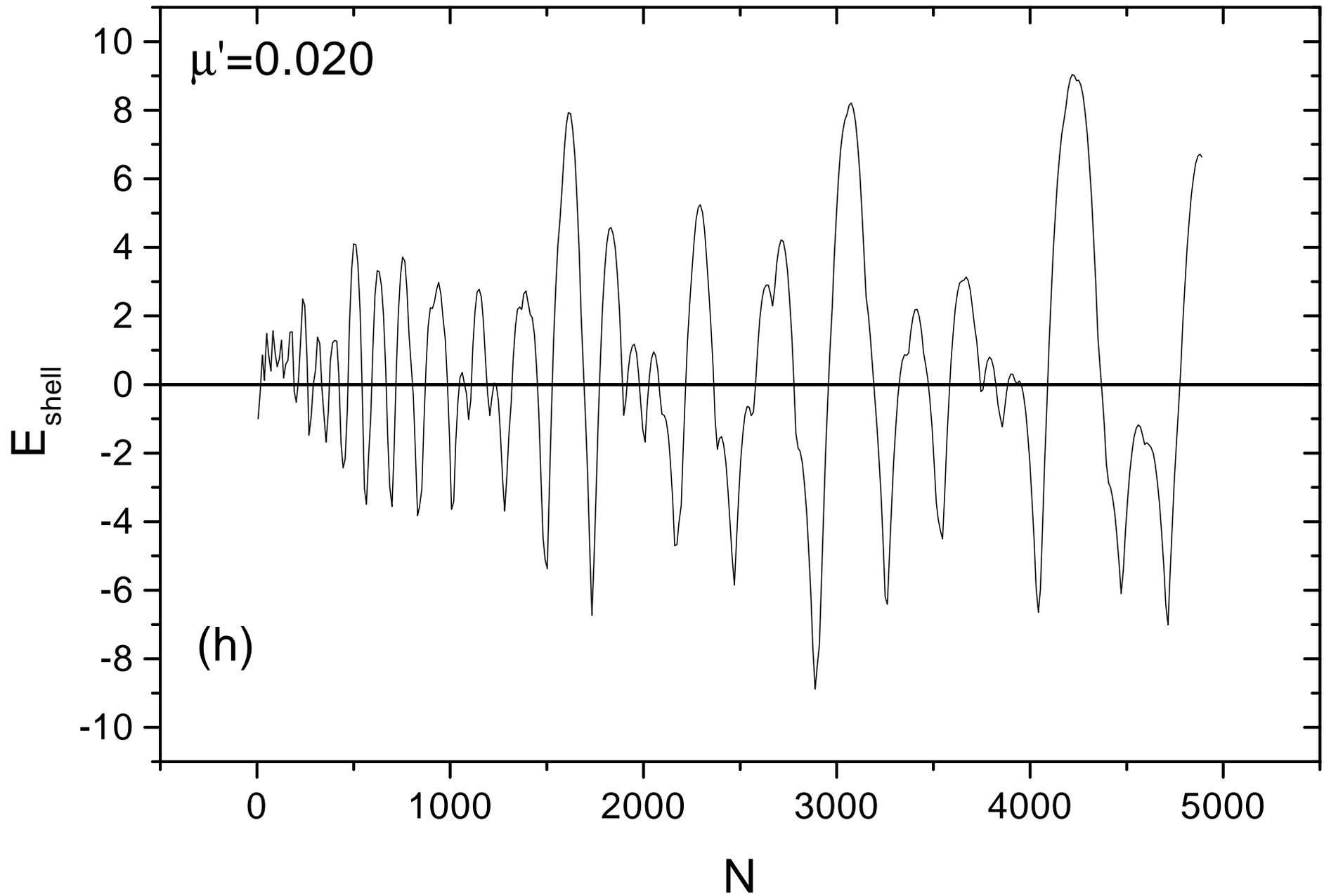
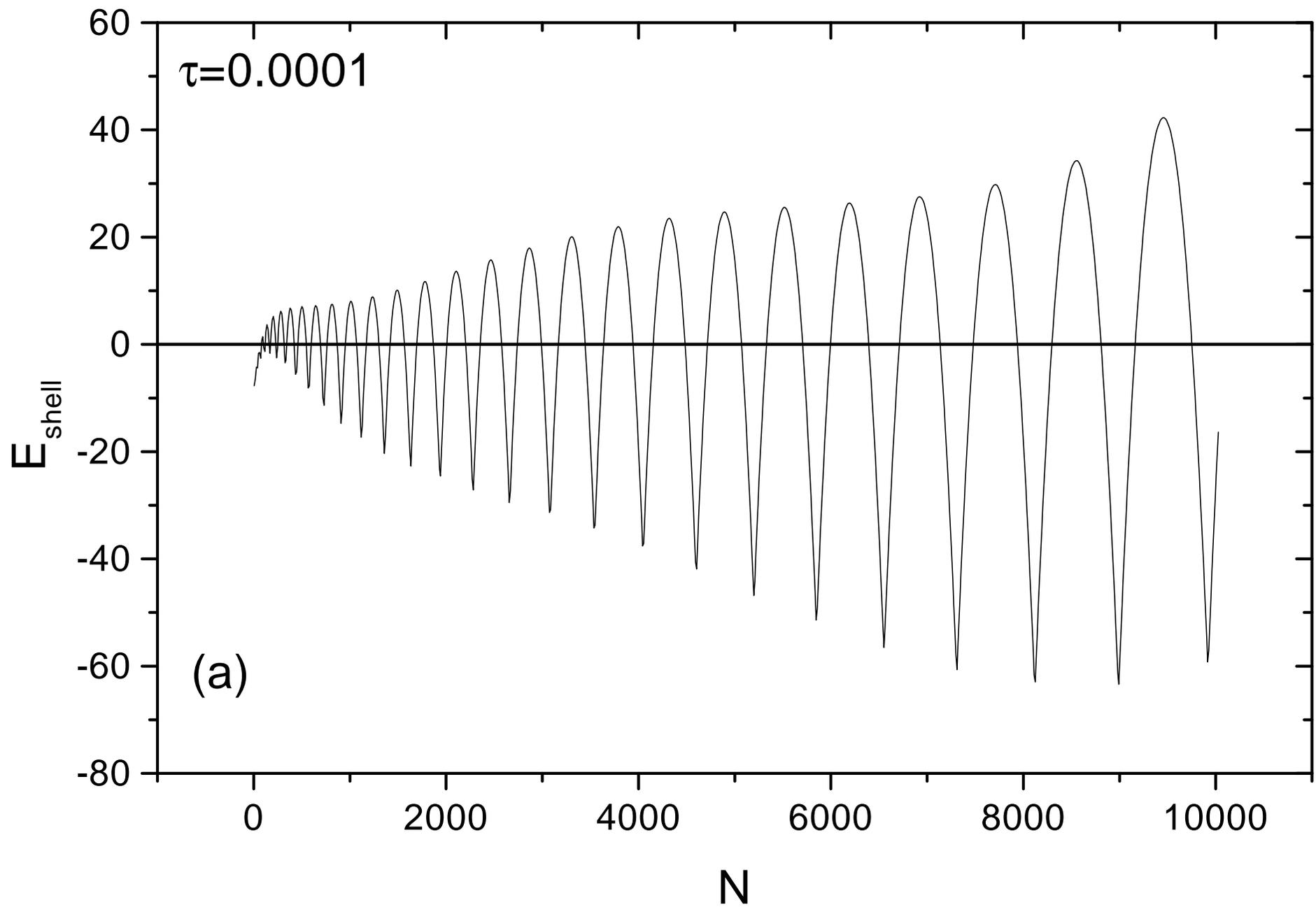
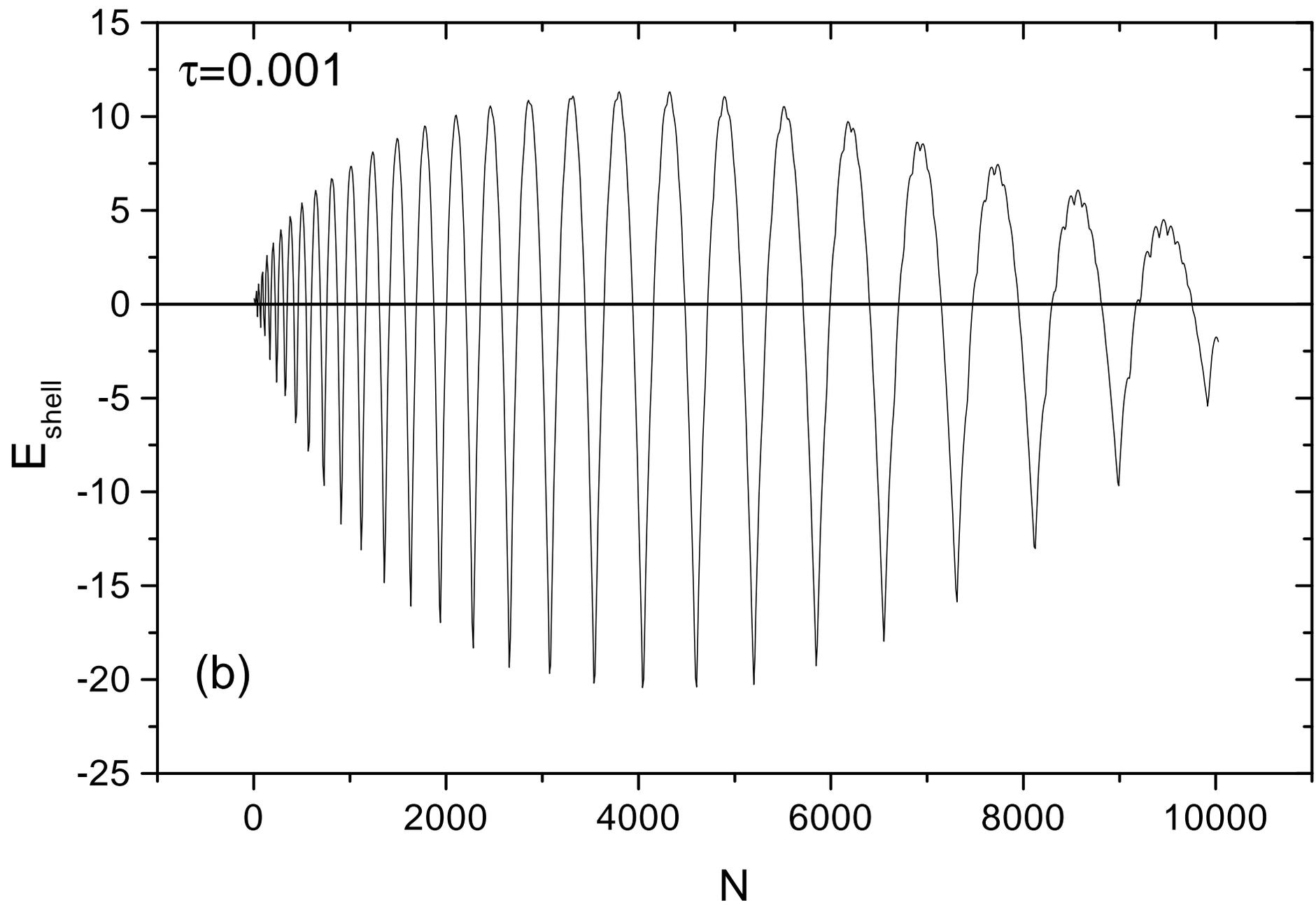
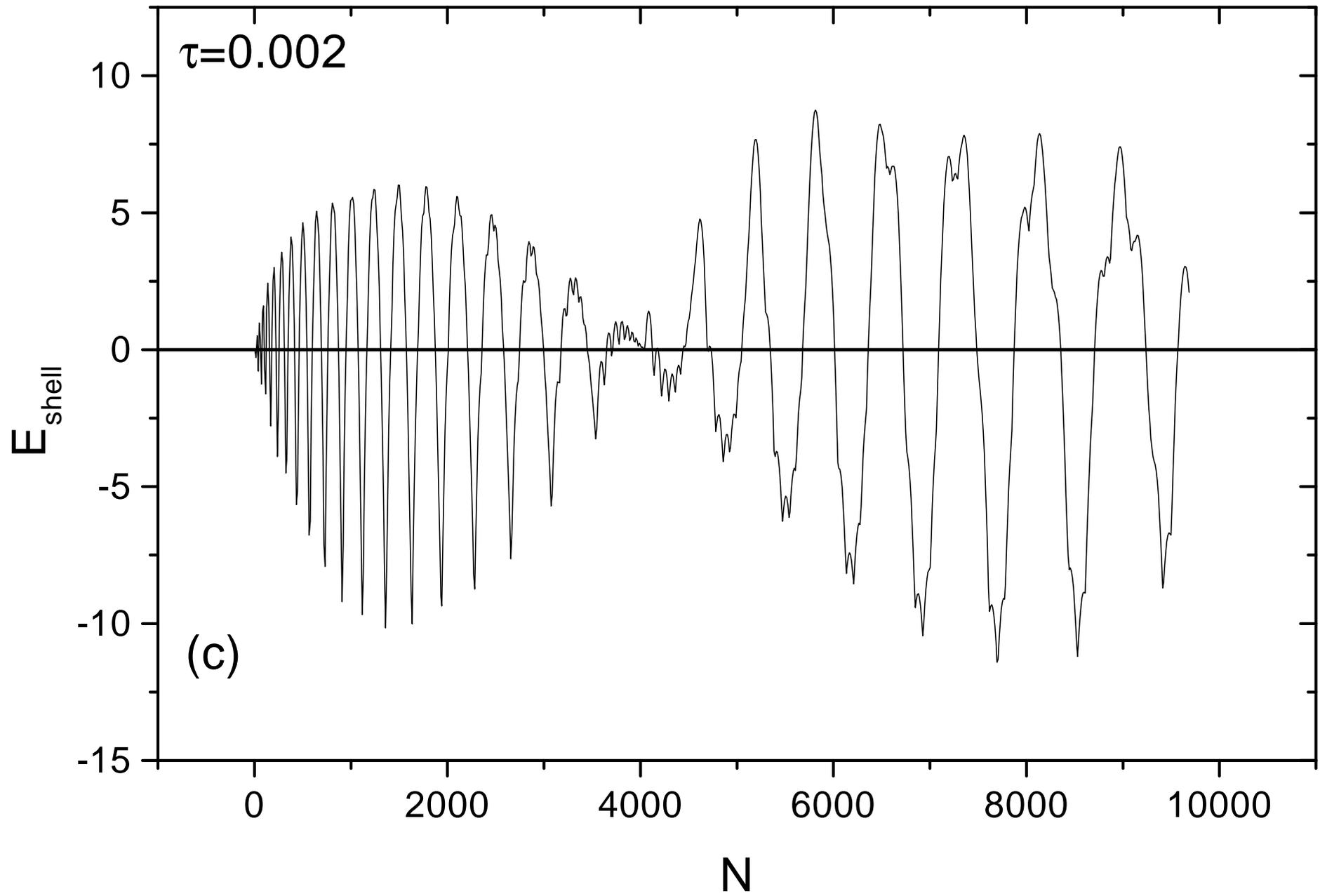
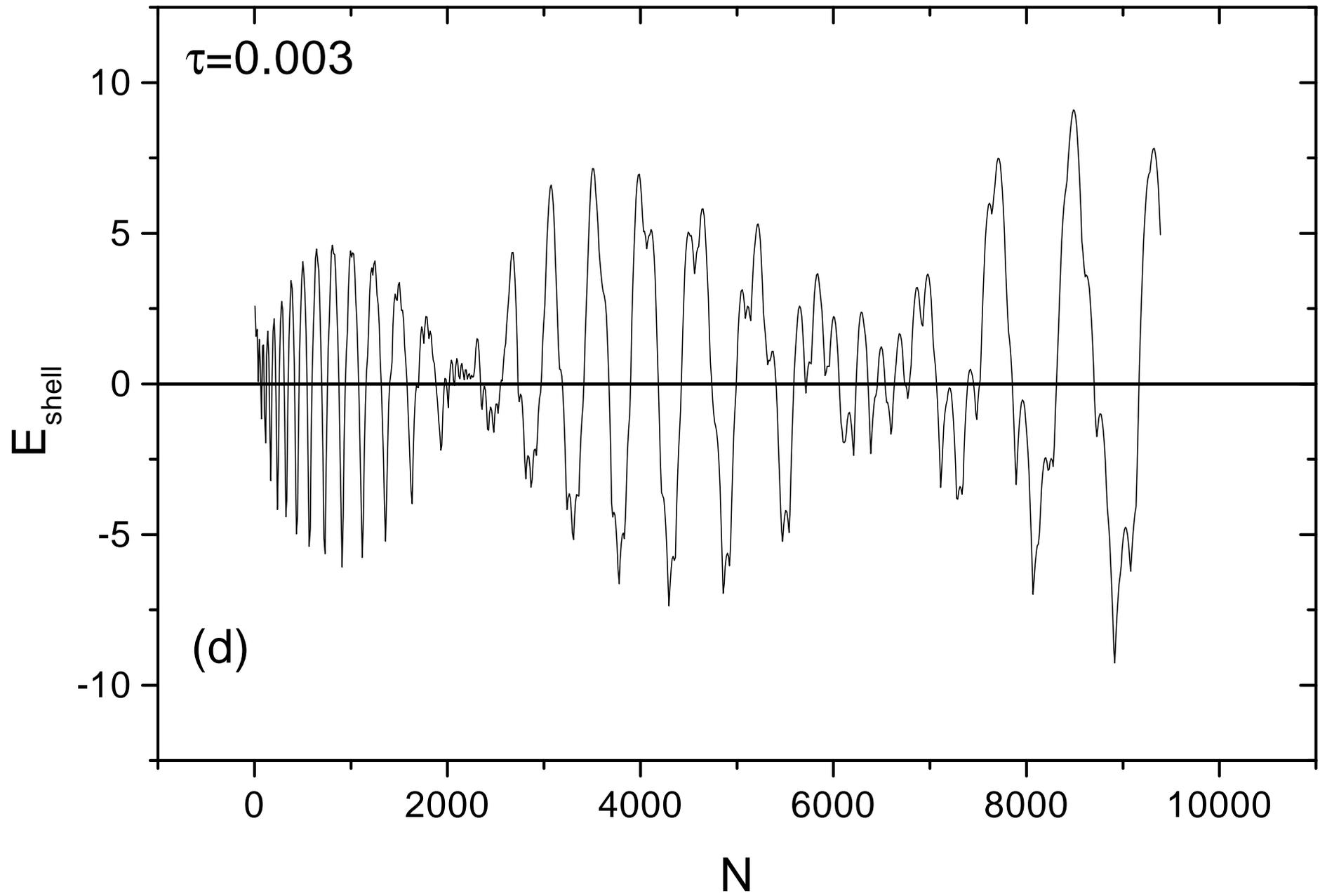
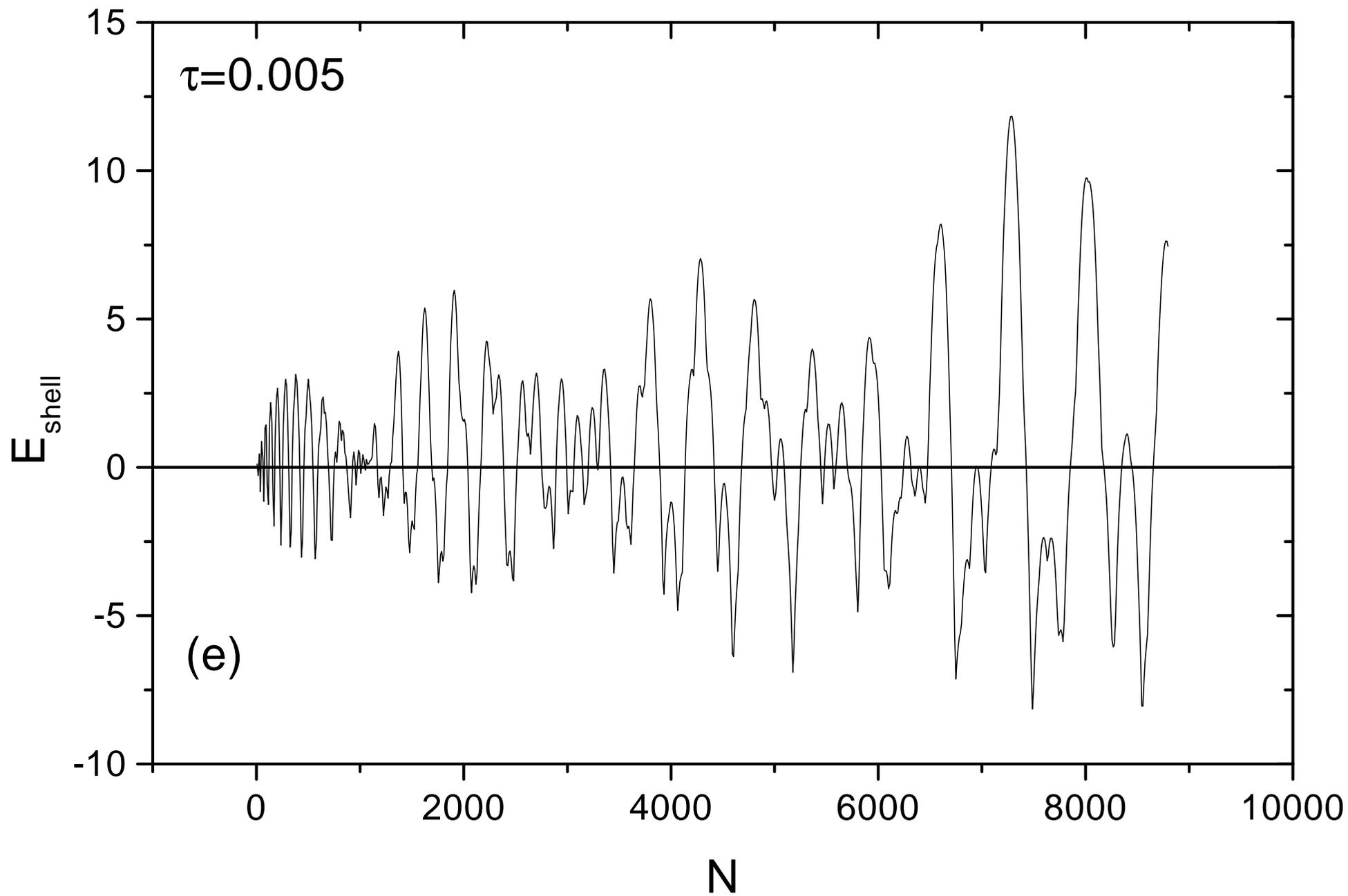
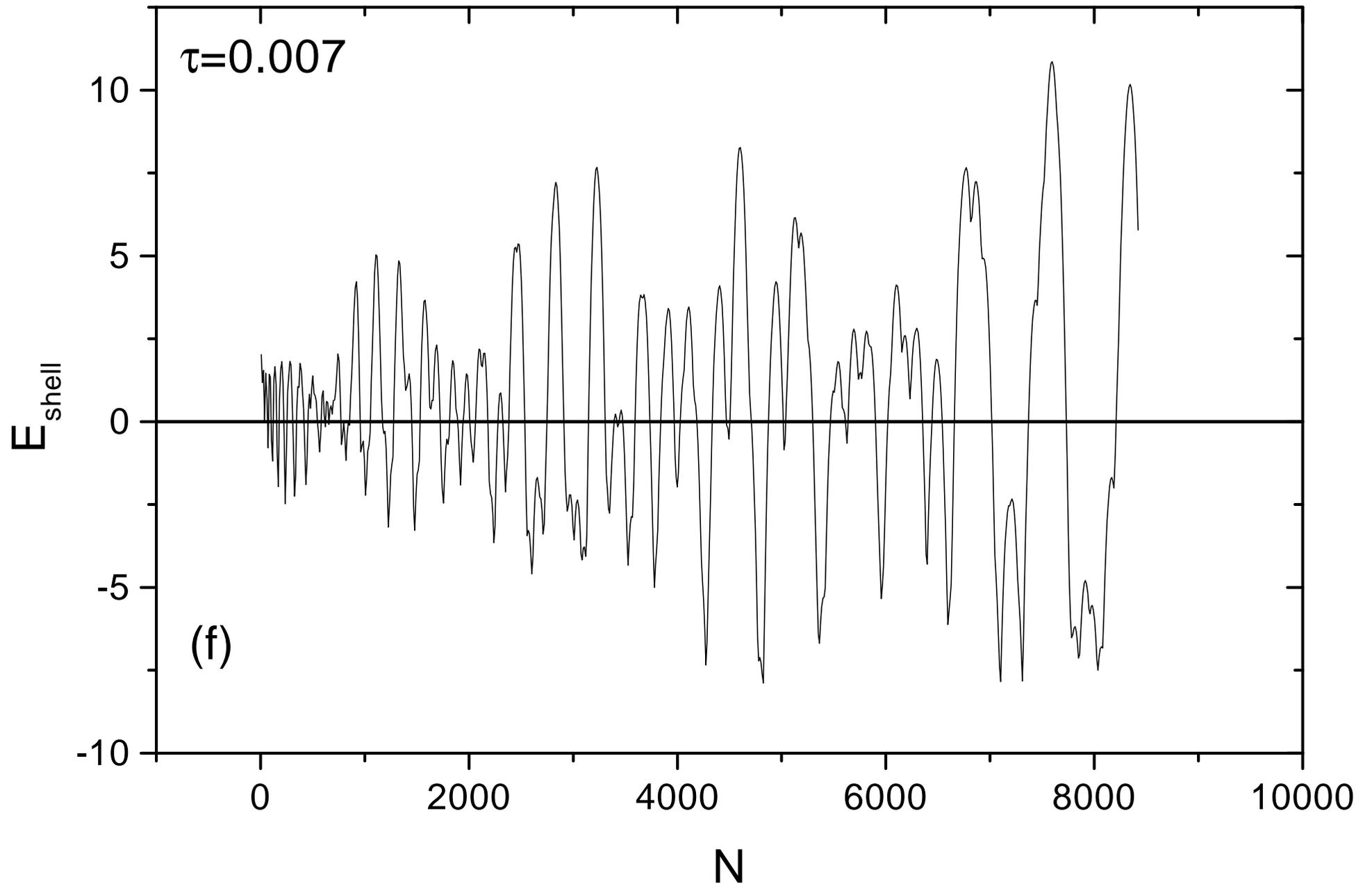
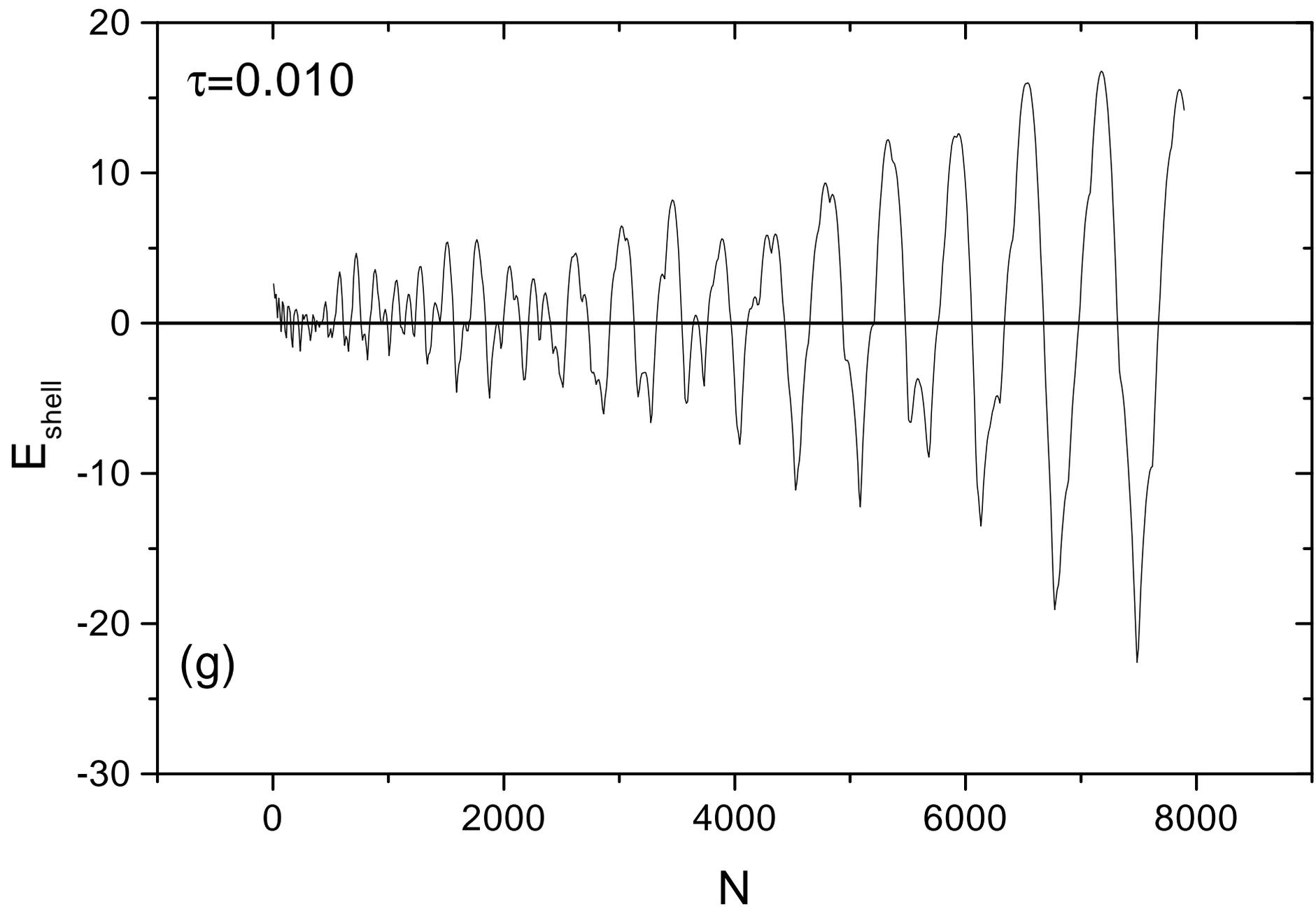
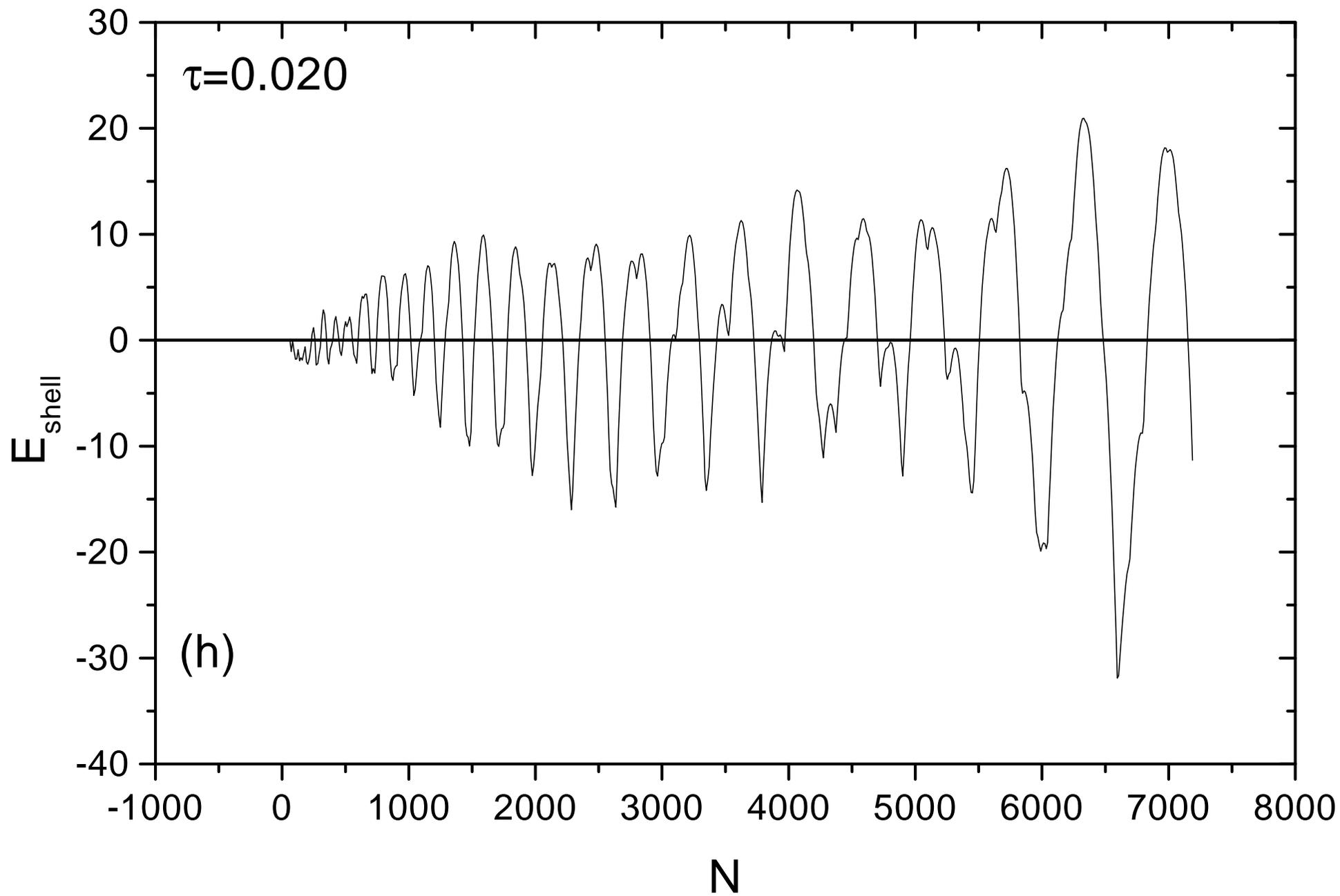
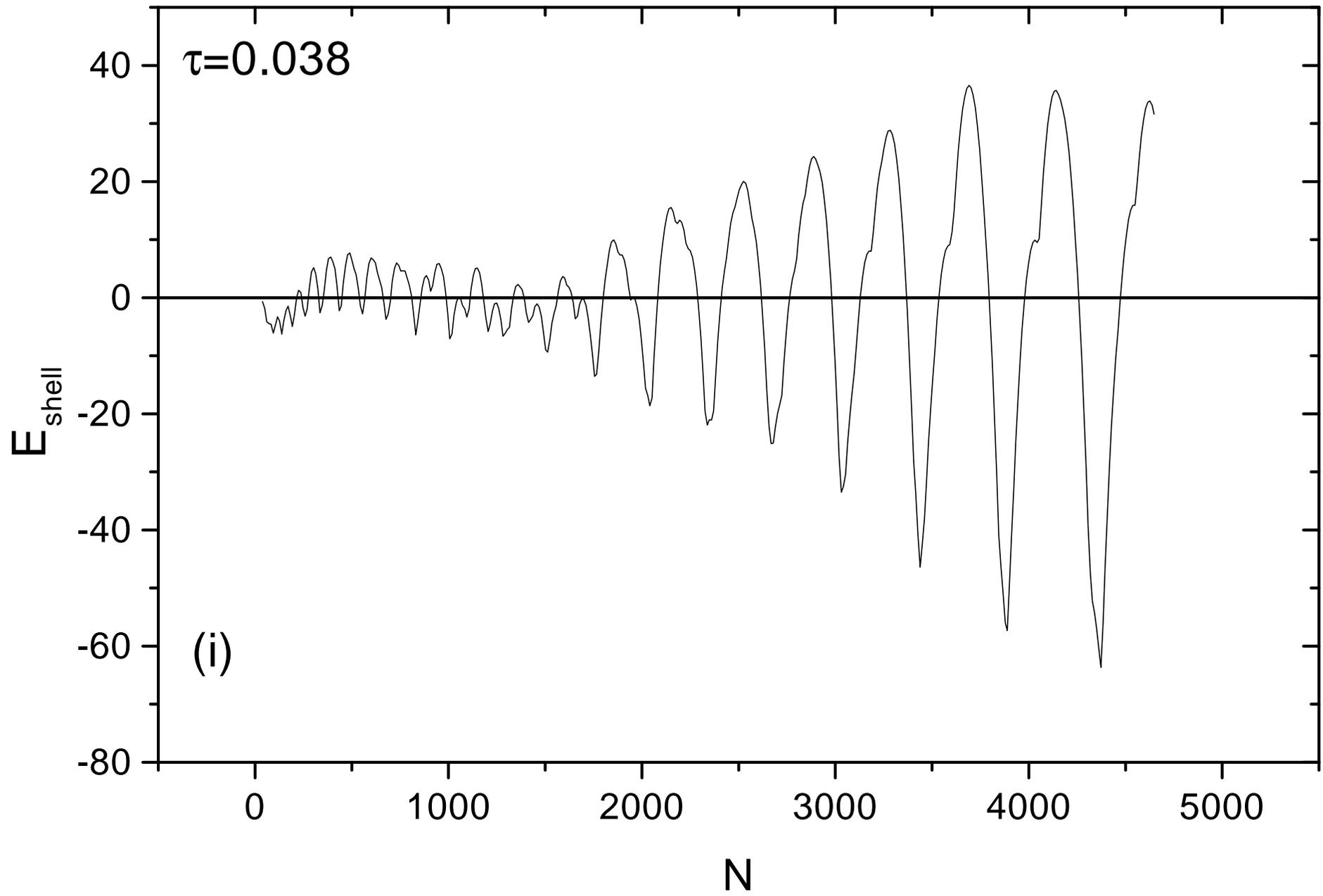
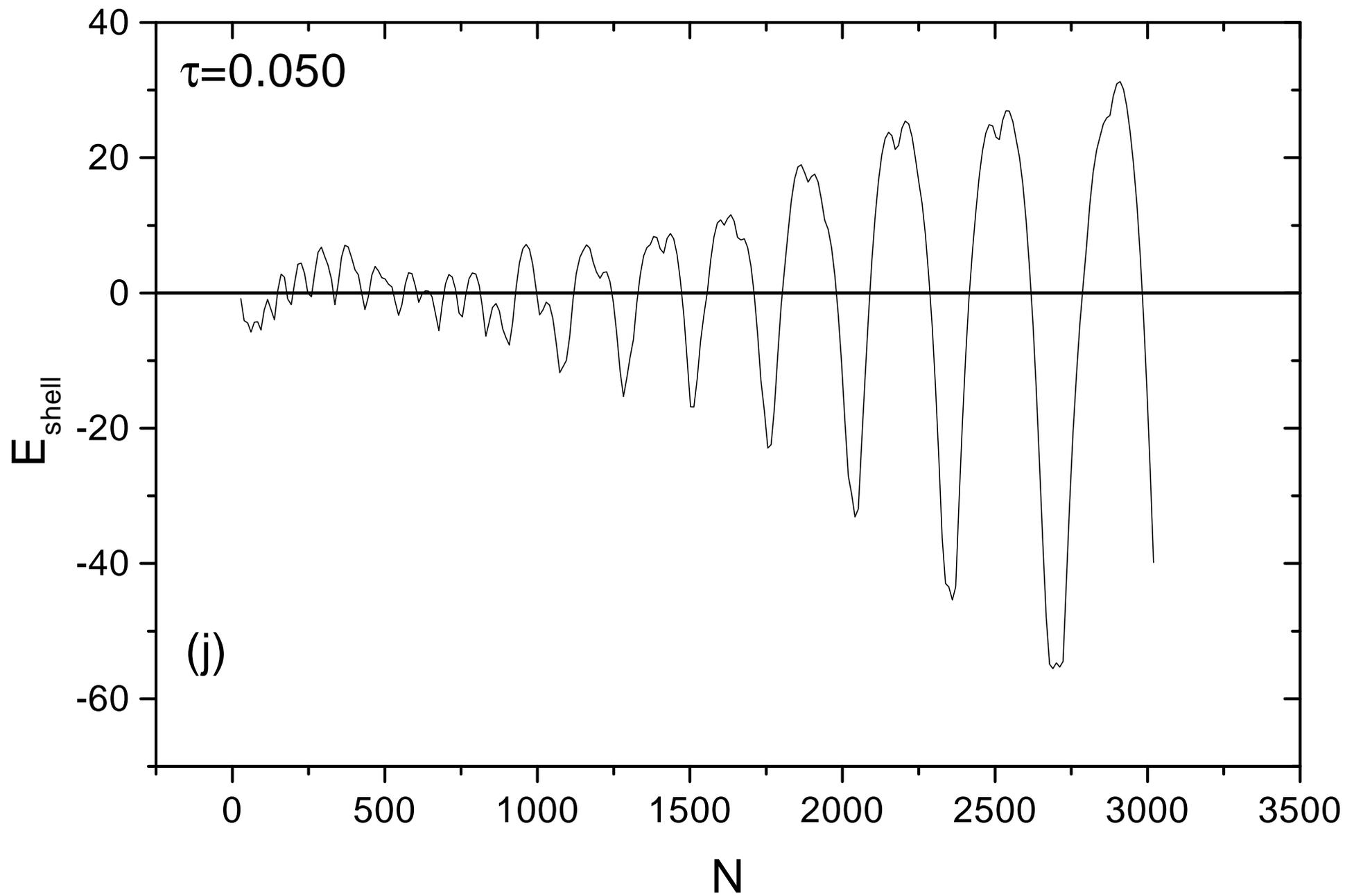
\end{document}